\documentclass[reprint,groupedaddress,showpacs,showkeys]{revtex4-2}
\usepackage{graphicx}
\usepackage{dcolumn}
\usepackage{epsfig}
\usepackage{xfrac}
\usepackage[version=3]{mhchem}
\usepackage{xcolor}
\usepackage{amsmath,amsfonts,amssymb}

\newbox\mybox
\newlength{\dzero}
\newcommand{\spcn}[1]{\settowidth{\dzero}{0}\kern#1\dzero}
\newcommand{\minnumb}{\settowidth{\dzero}{$-$}\kern-\dzero$-$}
\newcommand{\mystrut}{\rule[-0.5ex]{0.0em}{2.9ex}}
\begin{document}
\title[Electron spin-torsion coupling]{Evidence for electron spin-torsion coupling\\in
the rotational spectrum of the \ce{CH3CO} radical}
\author{Laurent H. Coudert}%
\email{Corresponding author: laurent.coudert@cnrs.fr}
\author{Olivier Pirali}
\author{Marie-Aline Martin-Drumel}
\author{Rosemonde Chahbazian}
\affiliation{Institut des Sciences Mol\'eculaires d'Orsay, Universit\'e Paris Saclay, CNRS, Orsay, France}
\author{Luyao Zou}
\altaffiliation{Present address: Universit\'e du Littoral C\^ote d’Opale, Laboratoire de Physico-Chimie de l'Atmosph\`ere,
UR4493, MREI2, 189A avenue Maurice Schumann, 59140 Dunkerque, France}
\author{Roman A. Motiyenko}
\author{Laurent Margul\`es} 
\affiliation{University of Lille, CNRS, UMR8523 - PhLAM - Physique des Lasers Atomes et Mol\'ecules, Lille, France}
\date{\today}
\begin{abstract}
Open-shell non-rigid molecular systems exhibiting an internal
rotation are likely candidates for a coupling between the spin
angular momentum of the unpaired electron and the torsional
motion.  This electron spin-torsion coupling lacked both an
experimental validation and a theoretical modeling. Here, the
first experimental observation of the electron spin-torsion
coupling is reported analyzing the pure rotational spectrum
at millimeter wavelengths of the \ce{CH3CO} radical,
a $^2\Sigma$ open-shell molecule displaying an internal
rotation of its methyl group. To account for this coupling,
a specific Hamiltonian incorporating the rotational, torsional,
and electronic degrees of freedom is developed and allows us to
reproduce the experimental spectrum. The present demonstration
of the electron spin-torsion coupling will undoubtedly be
key to future investigations of large open-shell molecules
exhibiting a complex internal dynamics.\end{abstract}

\maketitle

Non-rigid molecular systems have always challenged molecular
spectroscopists.  For instance, the internal rotation
of a methyl group, first theoretically investigated by
Nielsen~\cite{nielsen32}, was only satisfactorily modeled 30
years later~\cite{gebbie63, lees68} when the effects of the
strong Coriolis coupling between the overall rotation and the
torsional motion were fully understood.  While internal rotation
effects are nowadays well understood for close-shell molecules,
the main question for open-shell molecules concerns the coupling
of the unpaired electron spin with the torsional motion.
Its nature is still unknown and its effects have not been
evidenced yet.  The acetyl radical (\ce{CH3CO}) is a benchmark
system to study this  coupling as it displays a fine structure
and a large amplitude torsional motion of its methyl group
giving rise to the energy level diagram shown in
Ref.~\cite{supplement}
\nocite{lin59, bethe57, curl65, wilson_decius_cross, xu08, A79}%
and to sizable splittings observed in the pioneering
spectroscopic investigation of Hirota {\em et al.}~\cite{hirota07}.
However, the electron spin-torsion coupling could not have been
evidenced in the pure rotational $K_a=0$ transitions reported by
these authors because its effects vanish for this $K_a$-value,
as discussed below.

Observing the electron spin-torsion coupling in
the \ce{CH3CO} radical requires more extensive
laboratory measurements involving higher
$K_a$-values.  We recorded the pure rotational spectrum of
\ce{CH3CO} at millimeter and submillimeter wavelengths
(140--660~GHz) using our recently developed Zeeman-modulation
spectrometer~\cite{chahbazian24}. The radical was synthesized
{\em in situ} by hydrogen abstraction of acetaldehyde precursor
using fluorine atoms, a technique especially efficient when
investigating relatively large radicals as demonstrated
for the venoxy (\ce{CH2CHO}) and hydroxymethyl (\ce{CH2OH})
radicals~\cite{A104, chahbazian24}.
275 transitions with well resolved torsional and fine splittings
could be measured up to $N=22$ and $K_a=4$.
96 transitions with a resolved hyperfine structure are
also available including those previously recorded at centimeter
wavelengths~\cite{hirota07} and 18 new $K_a=0$ transitions recorded
in this work up to $N=5$ at millimeter wavelengths (75--100~GHz) using a
supersonic jet electrical discharge apparatus with acetaldehyde
and chlorine as precursors~\cite{zou2020, gyawali2023}.  The value
of the rotational quantum number $K_a$, larger than in Ref.~\cite{hirota07}, reached
in this study, together with a physical
derivation of the electron spin-torsion coupling, allows for
its first unambiguous observation in the \ce{CH3CO} radical
and is the subject of the present letter.

\begin{figure}
\centering
\includegraphics[width=0.9\linewidth]{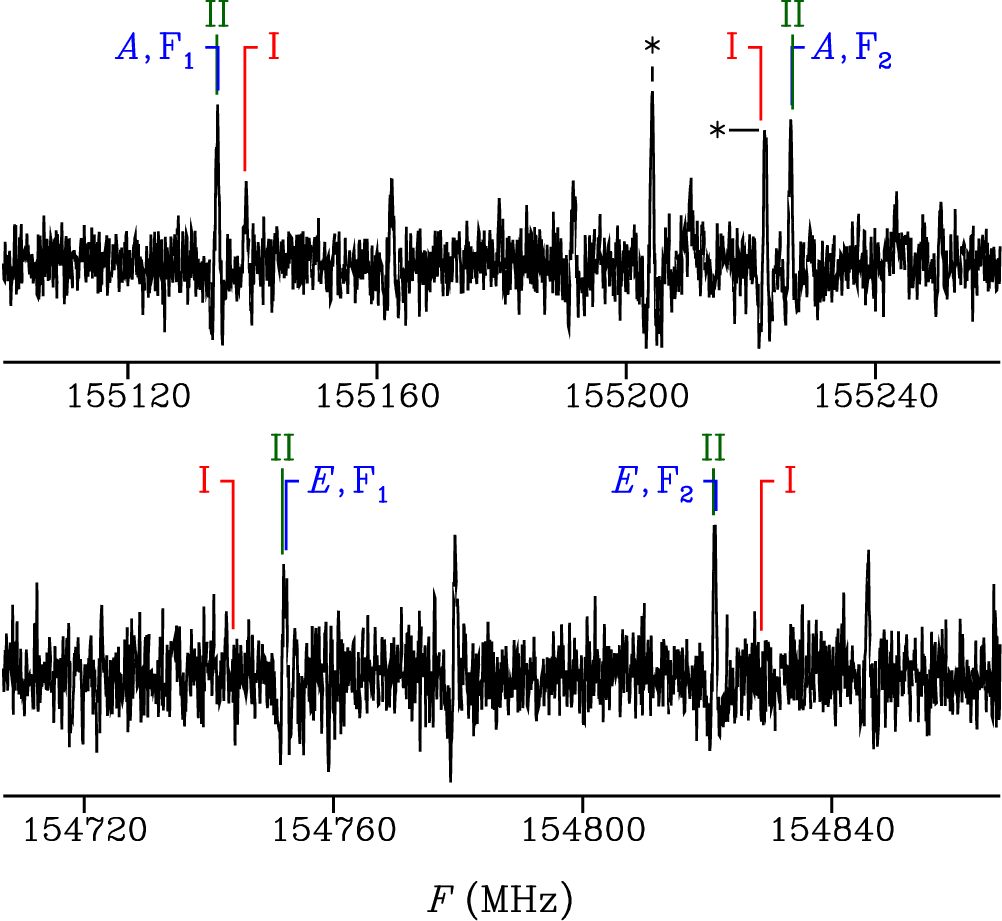}
\caption{\label{spectrum}Two portions of the observed millimeter
wave spectrum, as a function of the frequency in MHz, where
the four fine structure/torsional components of the $8_{26}
\leftarrow 7_{25}$ transition are observed. Each component is
assigned in terms of the $A$ and $E$ torsional symmetries and
the F$_1$ and F$_2$ fine splitting labels.  Their calculated
frequency with Model~I (II) is indicated by a vertical line with
an I (II). For both $A$ components, the match of the Model~I
frequencies with observed lines is fortuitous.  Lines with an
asterisk have been assigned to other CH$_3$CO transitions, the
remaining ones are unassigned to date.  The spectral analysis
is secured by the assignment of multiple series of lines
(not represented in this figure) performed using standard
Loomis-Wood plots~\cite{Bonah2022}.}\end{figure}

Line frequency analyses of the experimental data measured
in this work and in Ref.~\cite{hirota07}
were carried out with two theoretical models, referred to
as Models~I and II, where the electron spin-torsion coupling
was ignored and taken into account, respectively. The Hamiltonian used in
Model~I is the sum $H_{\rm RAM} + H_{\rm SR}$ of two model
Hamiltonians.  $H_{\rm RAM}$, based on previous theoretical
results~\cite{hecht57, lin59, kirtman62, lees68, herbst84},
is widely used in spectroscopic investigations of molecules
undergoing internal rotation~\cite{xu08} and is written using
the $\rho$ axis method (RAM)~\cite{hougen94}.  $H_{\rm SR}$ is
the Hamiltonian introduced by Brown and Sears~\cite{brown79}
to describe the electron spin-rotation coupling and includes
distortion effects. With a total of 32 spectroscopic parameters,
Model~I led to unsatisfactory results as the root
mean square (RMS) deviations of the observed minus calculated
frequencies far exceeded the experimental uncertainty of the
two data sets.  For the 275 millimeter and sub-millimeter
wave transitions, the RMS value was 2.27~MHz which is more than 22 times
their experimental uncertainty of 100~kHz; for the 96 
transitions with resolved hyperfine structure, 
the RMS value was 0.89~MHz which is 18 times their
experimental uncertainty of 50~kHz.  Model~I fails
to accurately compute the fine splittings of the two torsional
components of the observed lines. This can clearly be seen in the spectrum
depicted in Fig.~\ref{spectrum} where
the calculated fine splitting is
underestimated (overestimated) for the $A$-type ($E$-type)
torsional component.

Model~II includes the electron spin-torsion coupling which
is theoretically derived using the approach developed by
Curl~\cite{curl65} for treating fine effects in $^2\Sigma$
open-shell polyatomic molecules.  A molecule exhibiting an
internal rotation of its methyl group, parameterized by the
angle $\gamma$, is considered.  The molecule fixed position
vectors of the $n_n$ nuclei ${\bf a}_{\alpha}(\gamma)$
correspond to the RAM~\cite{hougen94} and the nuclear kinetic
energy is expressed with the inertia tensor ${\bf I}_n$,
with the direction cosines $\lambda_x,\lambda_y,\lambda_z$
of the axis of internal rotation, and with $I^{\alpha}$
the methyl group moment of inertia along its $C_3$ symmetry
axis~\cite{lin59, hougen94}. Describing the $n_e$ electrons by
their molecule fixed position vectors ${\bf r}_i$ and dropping
the translation, the total kinetic energy can be written:
\begin{equation}
\begin{split}
2 T&= \boldsymbol{\omega} \cdot {\bf I}_e \cdot \boldsymbol{\omega}
+ 2 \boldsymbol{\omega} \cdot \sum_i m ({\bf r}_i \times \dot{\bf r}_i)
+ \sum_i m \dot{\bf r}_i^2\\[1.0ex]
&+ \boldsymbol{\omega} \cdot {\bf I}_n \cdot \boldsymbol{\omega}
+ 2 \dot{\gamma} \, \boldsymbol{\omega} \cdot {\bf C}_{\gamma} + \dot{\gamma}^2 I^{\alpha}
\end{split}\label{ecine}
\end{equation}
where ${\bf I}_e$ is the inertia tensor of the electrons;
$\boldsymbol{\omega}$ is the angular velocity vector; $m$
is the mass of the electron; and ${\bf C}_{\gamma}$ is the
vector $\boldsymbol{\lambda} I^{\alpha}$.  The Lagrangian
$\mathcal{L}$ including electron spin couplings~\cite{curl65}
derived from Eq.~\eqref{ecine} can be written as:
%
%
\begin{equation}
\begin{split}
\mathcal{L}&= T +\boldsymbol{\omega}\cdot {\bf S}_{\rm R} %
+ \boldsymbol{\omega}\cdot \sum_i (-e/c) {\bf r}_i \times {\bf A}_i + \dot{\gamma} S_{\gamma}\vspace{1.0ex}\\
&-\boldsymbol{\omega}\cdot \sum_i (\mu_{\rm B}/c) {\bf r}_i \times ({\bf S}_i \times \boldsymbol{\mathcal{E}}_i)\vspace{1.0ex}\\
&+\sum_i (-e/c) {\bf A}_i\cdot \dot{\bf r}_i
-\sum_i (\mu_{\rm B}/c) ({\bf S}_i\times\boldsymbol{\mathcal{E}}_i) \cdot \dot{\bf r}_i \vspace{1.0ex}\\
&+(\mu_{\rm B}^2/2mc^2) \sum_i ({\bf S}_i\times\boldsymbol{\mathcal{E}}_i)^2 - V,
\end{split}\label{lagrang}
\end{equation}
where $e$, $\mu_{\rm B}$, and $c$ are the usual fundamental physics
constants; ${\bf S}_i$ is the spin angular momentum of electron $i$; $V$ is
the potential energy function; and:
\begin{equation}
\begin{array}{l}
\displaystyle
{\bf S}_{\rm R} = \sum_{\alpha} (Z_{\alpha} e/c) ({\bf a}_{\alpha}(\gamma)\times {\bf A}_{\alpha}),\vspace{1.0ex}\\
\displaystyle
S_{\gamma} =  \sum_{\alpha} (Z_{\alpha} e/c) {\bf A}_{\alpha} \cdot \frac{{\rm d} {\bf a}_{\alpha}(\gamma)}{{\rm d}\gamma}.
\end{array}\label{quantities}
\end{equation}
In Eqs.~\eqref{lagrang} and \eqref{quantities},
the potential vectors ${\bf A}_i$ and ${\bf A}_{\alpha}$, and the electric field
$\boldsymbol{\mathcal{E}}_i$ are to be obtained from
Eqs.~(10a), (10b), and (10c) of Ref.~\cite{curl65}, respectively.
The classical Hamiltonian $H$ can now be written:
\begin{equation}
\begin{split}
H & ={\textstyle\frac{1}{2}}({\bf N} - {\bf L} - p_{\gamma} \boldsymbol{\lambda} + S_{\gamma} \boldsymbol{\lambda} - {\bf S}_{\rm R}) \cdot\boldsymbol{\mu}\vspace{1.0ex}\\
& \cdot ({\bf N} - {\bf L} - p_{\gamma} \boldsymbol{\lambda} + S_{\gamma} \boldsymbol{\lambda} - {\bf S}_{\rm R})
+\frac{1}{2I^{\alpha}} (p_{\gamma} - S_{\gamma})^2 \vspace{1.0ex}\\
& + \frac{1}{2m} \sum_i [{\bf p}_i + (e/c) {\bf A}_i + (\mu_{\rm B} /c) ({\bf S}_i \times \boldsymbol{\mathcal{E}}_i)]^2\vspace{1.0ex}\\
& -(\mu_{\rm B}^2/2mc^2) \sum_i ({\bf S}_i \times \boldsymbol{\mathcal{E}}_i)^2 + V,
\end{split}\label{hamilt}
\end{equation}
where ${\bf N}$ is the rotational angular momentum conjugate to
$\boldsymbol{\omega}$; $p_{\gamma}$ is the momentum conjugate
to $\dot{\gamma}$; ${\bf L}$ and ${\bf p}_i$ are defined as
for Eq.~(22) of Curl~\cite{curl65}; and $\boldsymbol{\mu}$
is a $3\times 3$ tensor equal to the inverse of ${\bf I}_n -
\boldsymbol{\lambda} \boldsymbol{\lambda}^{\intercal} I^{\alpha}$.
This tensor being
constant, $H$ also is the quantum-mechanical Hamiltonian.
See Ref.~\cite{supplement} for a detailed
derivation of H.

The Hamiltonian in Eq.~\eqref{hamilt} can be written as the
sum $H_{\rm RT} + H_e + H_{\rm S}$, where $H_{\rm RT}$ and
$H_e$ are respectively the spin-independent rotation-torsion
and electronic Hamiltonians and $H_S$ contains the remaining
spin-dependent terms. The rotation-torsion Hamiltonian assumes
the following form:
\begin{equation}
\begin{split}
H_{\rm RT} & ={\textstyle\frac{1}{2}}({\bf N} - p_{\gamma} \boldsymbol{\lambda}) \cdot\boldsymbol{\mu}
\cdot ({\bf N} - p_{\gamma} \boldsymbol{\lambda})\\[1.0ex]
&+\frac{p_{\gamma}^2}{2I^{\alpha}} + V_{\rm T}(\gamma),
\end{split}\label{hrt}
\end{equation}
where $V_{\rm T}(\gamma)$ is the hindering potential. $H_{\rm RT}$
can be shown to be identical to the RAM Hamiltonian
in Eq.~(1) of Xu {\em et al.}~\cite{xu08}
using the results in Refs.~\cite{lin59, hougen94, supplement}.
The electronic Hamiltonian $H_e$ takes the simple form:
\begin{equation}
\sum_i \, {\bf p}_i^2/2m + V - V_{\rm T}(\gamma).
\end{equation}
Only terms giving rise to electron spin-rotation or torsion are retained
in the operator $H_S$ which reduces to:
\begin{equation}
H^{(1)} + H_a^{(2)} + H_b^{(2)}.
\label{sumop}
\end{equation}
These three terms, listed below, should be symmetrized:
\begin{equation}
H^{(1)} = -({\bf N} - p_{\gamma} \boldsymbol{\lambda}) \cdot \boldsymbol{\mu}\cdot {\bf S}_{\rm R}
-2F(p_{\gamma} - \boldsymbol{\rho} \cdot {\bf N}) S_{\gamma}
\label{first_order}
\end{equation}
where $F$ and $\rho$ are defined in Refs.~\cite{hougen94, xu08}, gives
rise to a first order contribution.  The remaining terms:
\begin{equation}
\begin{array}{l}
\displaystyle
H_a^{(2)}= \sum_i \, [(e/mc) {\bf A}_i + (\mu_{\rm B}/mc) ({\bf S}_i \times \boldsymbol{\mathcal{E}}_i)]
\cdot {\bf p}_i,\vspace{1.0ex}\\
\displaystyle
H_b^{(2)}=-{\bf L}\cdot \boldsymbol{\mu}\cdot ({\bf N} - p_{\gamma} \boldsymbol{\lambda}),
\end{array}
\label{sec_order}
\end{equation}
lead to a second order contribution.
The first order effects of $H^{(1)}$ 
are accounted for as
in Refs.~\cite{curl65, brown80} and the following
effective Hamiltonian arises:
\begin{equation}
\begin{array}{l}
H_{\rm SRT}^{(1)}={\textstyle\frac{1}{2}}[{\bf S}\cdot \boldsymbol{\epsilon}^{(1)} \cdot
({\bf N} -p_{\gamma} \boldsymbol{\lambda})
+ ({\bf N} -p_{\gamma} \boldsymbol{\lambda}) \vspace{1.0ex}\\
\mbox{}\hspace{2.0em}  \cdot \boldsymbol{\epsilon}^{(1)} \cdot {\bf S}]
+ {\textstyle\frac{1}{2}}\{({\bf S} \cdot \boldsymbol{\epsilon}^{(1)}_{\gamma}),
(p_{\gamma} - \boldsymbol{\rho} \cdot {\bf N})\},\end{array}
\label{SRT1}
\end{equation}
where ${\bf S}$ is the effective operator corresponding to the
spin of the unpaired electron, $\boldsymbol{\epsilon}^{(1)}$
is the first order electron spin-rotation coupling tensor
to be retrieved from Eq.~(26) of Curl~\cite{curl65}, and
$\boldsymbol{\epsilon}^{(1)}_{\gamma}$ is a coupling vector
which can be evaluated as in this reference.  Using the same
procedure as in Section~1 of Curl~\cite{curl65}, the second
order effects of the two operators in Eqs.~\eqref{sec_order}
are evaluated leading to the effective Hamiltonian:
\begin{equation}
H_{\rm SRT}^{(2)} ={\textstyle\frac{1}{2}}
[{\bf S}\cdot \boldsymbol{\epsilon}^{(2)} \cdot ({\bf N} - p_{\gamma} \boldsymbol{\lambda})
+ ({\bf N} -p_{\gamma}\boldsymbol{\lambda})\cdot \boldsymbol{\epsilon}^{(2)} \cdot {\bf S}],
\label{SRT2}
\end{equation}
where ${\bf S}$ is defined as for Eq.~\eqref{SRT1}
and $\boldsymbol{\epsilon}^{(2)}$ is the second
order electron spin-rotation coupling tensor given in
Eq.~(6a) of Curl~\cite{curl65}.
As the second order contribution usually dominates the
first order one~\cite{curl65, brown77}, the effective
coupling Hamiltonian in Eq.~\eqref{SRT2} will be used
henceforth.  In this equation, $\frac{1}{2}({\bf S}\cdot
\boldsymbol{\epsilon}^{(2)} \cdot {\bf N}
+ {\bf N} \cdot \boldsymbol{\epsilon}^{(2)} \cdot {\bf S})$ is the usual
electron spin-rotation coupling and $-\frac{1}{2}({\bf
S}\cdot \boldsymbol{\epsilon}^{(2)} \cdot \boldsymbol{\lambda} \,p_{\gamma}
 + p_{\gamma} \boldsymbol {\lambda}\cdot \boldsymbol{\epsilon}^{(2)} \cdot {\bf S})$
is the sought electron spin-torsion operator.

Model~II leads to an improved model Hamiltonian expressed as
the sum $H_{\rm RAM} + H_{\rm SR} + H_{\rm ST}$ where
the last term, describing the electron spin-torsion
coupling, is written in agreement with
Eq.~\eqref{SRT2} as:
\begin{equation}
H_{\rm ST} = (e_x S_x + e_z S_z)  p_{\gamma},
\label{STfit}
\end{equation}
where $e_x$ and $e_z$ are electron spin-torsion constants.
No term in $S_y p_{\gamma}$ arises for symmetry reasons~\cite{hougen94}.

\begin{figure}
\centering
\includegraphics[width=0.9\linewidth]{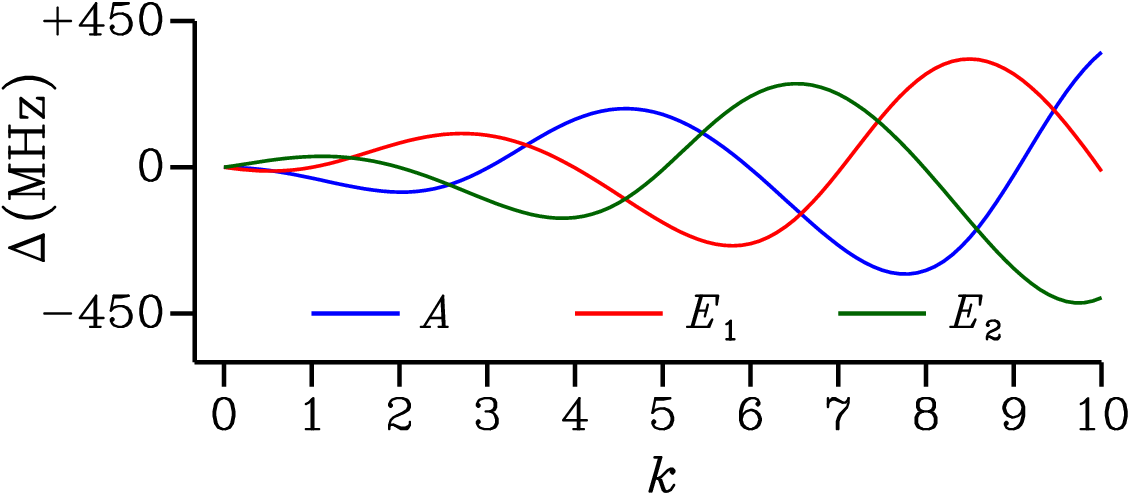}
\caption{\label{psi_A_plot}The fine splitting $\Delta=E({\rm
F}_1) - E({\rm F}_2)$ with Model~II is plotted in MHz as a function of the
rotational quantum number $k$ for $N=10$ and for the three
$v_t=0$ torsional levels identified by their color coded $C_3$
species $A$, $E_1$, $E_2$.}\end{figure}

The effects of the electron spin-torsion coupling are qualitatively
investigated with the help of a torsional Hamiltonian $H_{\rm
T}$ built from the improved model Hamiltonian by extracting
those operators with diagonal matrix elements in the coupled
$|NkSJ\rangle$ Hund's case {\em b} rotation-spin basis set
functions.  Discarding terms leading to a slow variation with
$N$ or $k$, we obtain:
\begin{equation}
H_{\rm T} = F (p_{\gamma} -\rho N_z)^2 + e_z S_z p_{\gamma} + e_z S_z N_z/2 + V_{\rm T}(\gamma),
\label{torsionalham}
\end{equation}
where $e_z S_z p_{\gamma}$ is the term in $H_{\rm ST}$
giving rise to a diagonal
contribution and the term $S_z N_z$, with no physical meaning,
removes an unwanted variation of the fine splitting.
The torsional Hamiltonian $H_{\rm T}$
is diagonalized using the free internal rotation basis
set functions $|m\rangle=\exp(im\gamma)/\sqrt{2\pi}$.
Torsional levels with $C_3$ symmetries $A$, $E_1$, and
$E_2$ arise respectively for $m=3p$, $3p+1$, and $3p-1$,
where $p$ is an integer~\cite{lin59, supplement}.  The constants in
Eq.~\eqref{torsionalham} were set to the values reported
in Table~\ref{params}.  The variations of
the fine splitting, defined as the energy difference $E({\rm
F}_1) - E({\rm F}_2)$, are plotted in Fig.~\ref{psi_A_plot}
as a function of $k$ for the three $C_3$ symmetries and for
$v_t=0$.  The electron spin-torsion leads to fine splittings
depending on the torsional symmetry of the levels as well
as on $k$, which is consistent with the experimental data recorded in this work.
It vanishes for $k=0$ which
is why its effects could not be observed in the $K_a=0$
transitions published previously~\cite{hirota07}.
Its semi-periodic variations are consistent
with the periodic variations of the torsional splittings determined in
the pioneering investigations of internal rotation~\cite{burkhard59}.
It scales as $1/N$ just like the electron spin-rotation coupling.

\begin{figure}
\centering
\includegraphics[width=0.9\linewidth]{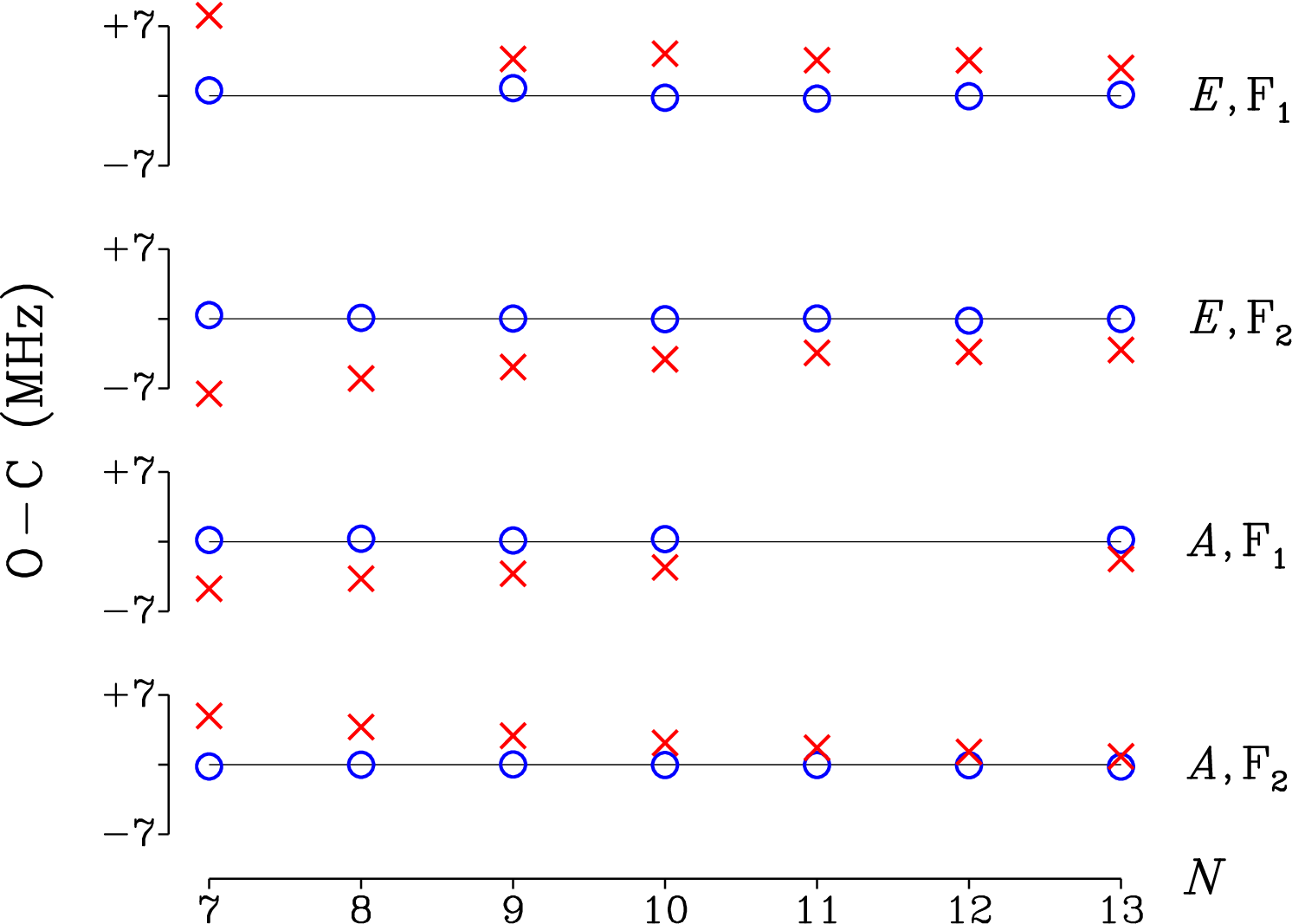}
\caption{\label{loomis_plot}For $R(N)$, $K_a=2$, $a$-type
transitions belonging to the upper component of the
asymmetry doublet, the observed minus calculated frequency
($\mbox{O}-\mbox{C}$) is plotted in MHz as a function of $N$
with multiplication and circle symbols for Models~I and II,
respectively.  Each panel corresponds to either $A$- or
$E$-type torsional sublevels and to either F$_1$ of F$_2$
fine components.  The height of the symbols is for clarity 25
times the experimental uncertainty, 100~kHz.}\end{figure}

In the analysis carried out with Model~II, a total of 37 parameters were
considered including the 32 parameters used in Model~I and 5 parameters specific to the electron spin-torsion
coupling. This second analysis is much more satisfactory than the first one.
For the 275 millimeter and sub-millimeter
wave transitions, the RMS value was 0.35~MHz which is less
than 4 times their experimental uncertainty; for the 96
transitions with resolved hyperfine structure, the RMS value was 0.13~MHz which is
less than 1.5 times their experimental uncertainty.  This improvement can clearly be
seen in Fig.~\ref{loomis_plot}, especially for small $N$ values, as expected from
the $1/N$ scaling of the electron spin-torsion coupling. Table~\ref{params} reports the
parameters determined in this second analysis excluding distortion and hyperfine
coupling parameters. The electron spin-torsion
coupling parameters $e_x$ and $e_z$ are well defined and display uncertainties
respectively 53 and 110 times smaller than their values.
Equation~\eqref{SRT2} allows us to express these parameters
in terms of components of the electron spin-rotation coupling
tensor:
\begin{equation}
e_{\beta}^{\rm cal} = -\epsilon_{\beta x} \lambda_x -\epsilon_{\beta z} \lambda_z, \mbox{  where  $\beta=x,z$.}
\end{equation}
Assuming $\epsilon_{xz}=\epsilon_{zx}$, in agreement with
Ref.~\cite{hirota07}, and determining the direction cosines
$\lambda_x$ and $\lambda_z$ from the structure given in this
reference leads to $e_x^{\rm cal}=-20.8(11)$ and $e_z^{\rm
cal}=-1992.3(97)$~MHz.  The agreement with
the value in Table~\ref{params} is unexpectedly good for the
well defined and larger $e_z$ as both values are within 2.3\%.

\begin{table}
\caption{\label{params}Spectroscopic constants from Model~II excluding
distortion and hyperfine coupling constants}
\centering
\begin{ruledtabular}
\begin{tabular}{@{}l l c@{}}
Name &
Operator\footnote{$S_{xz}^+$ is the operator $[\{S_x, N_z\} + \{S_z, N_x\}]/2$.} &
Value/MHz\footnote{$V_3$ is in cm$^{-1}$ and $\rho$ is unitless. Uncertainties in parentheses in the same units as the last quoted digit.} \\ \hline

\multicolumn{3}{@{}l}{Internal rotation\footnote{Defined in Eq.~(1) of Ref.~\cite{xu08}.}\mystrut} \\
 $V_3$            & $(1-\cos 3\gamma)/2$ & \spcn{0}144.72(15)\spcn{4} \\
 $F$              & $p_{\gamma}^2$       & \spcn{-3}296758.(189)\spcn{5} \\
 $\rho$           & $p_{\gamma} N_z$     & \spcn{2}\minnumb0.498963(11)\spcn{0} \\
 $A$              & $N_z^2$              & \spcn{-2}83904.3(1072)\spcn{3} \\
 $B$              & $N_x^2$              & \spcn{-2}10128.944(39)\spcn{3} \\
 $C$              & $N_y^2$              & \spcn{-1}9419.819(39)\spcn{3} \\
 $D_{xz}$         & $\{N_x, N_z\}$       & \spcn{-1}\minnumb3407.49(48)\spcn{4}\vspace{1.5ex} \\
\multicolumn{3}{@{}l}{Electron spin-rotation\footnote{Defined in Eq.~(1) of Ref.~\cite{brown79}.}\mystrut} \\
 $\epsilon_{xx}$  & $S_x N_x$            & \spcn{1}11.72(19)\spcn{4} \\
 $\epsilon_{yy}$  & $S_y N_y$            & \spcn{1}\minnumb43.28(18)\spcn{4} \\
 $\epsilon_{zz}$  & $S_z N_z$            & \spcn{-1}2096.6(99)\spcn{5} \\
 $\sfrac{1}{2} (\epsilon_{xz}^0 + \epsilon_{zx}^0)$ & $S_{xz}^+$           & \spcn{1}18.0(11)\spcn{5}\vspace{1.5ex} \\
\multicolumn{3}{@{}l}{Electron spin-torsion\footnote{Defined in Eq.~\eqref{STfit}.}\mystrut} \\
 $e_x$            & $p_{\gamma} S_x$     & \spcn{0}\minnumb152.2(28)\spcn{5} \\
 $e_z$            & $p_{\gamma} S_z$     & \spcn{-1}\minnumb2038.(19)\spcn{6} \\

\end{tabular}
\end{ruledtabular}

\end{table}

The results reported in this letter conclusively confirm that
a novel interaction, referred to as
the electron spin-torsion coupling, has been evidenced in the
\ce{CH3CO} radical and theoretically accounted for. This claim is supported by three results.
({\em i}) Accounting solely for the internal
rotation of the methyl group and the electron spin-rotation
coupling precludes accurately reproducing experimental fine splittings.
({\em ii}) A model Hamiltonian where the new coupling is
included allows us to successfully reproduce the high-resolution
spectroscopic data of the radical and specifically the fine
splittings.  ({\em iii}) The relation between the experimental
value of the electron spin-torsion constants and their
theoretical value, deduced from the electron spin-rotation
coupling tensor, is correct within 2.3\% for the largest and
best defined constant.

The electron spin-torsion coupling will be key to build the
spectroscopic databases to be used for astronomical searches
of the \ce{CH3CO} radical, suspected to be an important
intermediate to the formation of complex organic molecules
in space~\cite{Ferrero2023, ijms241914510}. An astrophysical detection is
quite likely as its cation \ce{CH3CO+} has already been
observed in the interstellar medium~\cite{cernicharo21}.
This detection will require catalogs with frequency accuracies of 1 MHz or lower.
It is obvious from the present study that neglecting the electron spin-torsion coupling
yields modeling errors of several MHz, Fig.~\ref{loomis_plot},
precluding an interstellar detection. More generally,
large radicals (6 atoms and more) are presently under high scrutiny owing to
their central role in interstellar chemistry~\cite{herbst2009, C6CP07024H}.
Future spectroscopic investigations of those containing a methyl group
will undoubtedly require considering the electron spin-torsion coupling.
The rotational spectrum of one such radical, the non-rigid acetonyl radical (\ce{CH3COCH2}),
has recently been recorded thanks to a new experimental technique~\cite{Chahbazian2024}.
Preliminary results obtained in our group indicate that the electron spin-torsion
coupling needs to be considered to model its spectrum.

%

The present theoretical treatment could be extended to molecules exhibiting internal rotation of a $C_{2v}$
symmetry group, for instance, to the non-rigid \ce{O2-H2O} complex. In this atmospherically relevant
species, one of the challenges will be
to model the electron spin-torsion in a triplet $^3\Sigma$ state~\cite{acp-11-8607-2011}.

The work at ISMO was performed thanks to financial support
from LabEx PALM (ANR-10-LABX-0039-PALM), from the R\'egion
Ile-de-France through DIM-ACAV+, and from the Agence Nationale
de la Recherche (ANR-19-CE30-0017-01).  We acknowledge support
from the Programme National ``Physique et Chimie du Milieu
Interstellaire'' (PCMI) of CNRS/INSU with INC/INP co-funded by
CEA and CNES.  L.\ Z.\ acknowledges support from the European
Union's Horizon 2020 research and innovation programme under
the Marie Sk\l{}odowska-Curie Individual Fellowship grant
(H2020-MSCA-IF-2019, Project no.\ 894508).

\bibliography{gbib,ch3co,supp}

\begin{thebibliography}{31}%
\makeatletter
\providecommand \@ifxundefined [1]{%
 \@ifx{#1\undefined}
}%
\providecommand \@ifnum [1]{%
 \ifnum #1\expandafter \@firstoftwo
 \else \expandafter \@secondoftwo
 \fi
}%
\providecommand \@ifx [1]{%
 \ifx #1\expandafter \@firstoftwo
 \else \expandafter \@secondoftwo
 \fi
}%
\providecommand \natexlab [1]{#1}%
\providecommand \enquote  [1]{``#1''}%
\providecommand \bibnamefont  [1]{#1}%
\providecommand \bibfnamefont [1]{#1}%
\providecommand \citenamefont [1]{#1}%
\providecommand \href@noop [0]{\@secondoftwo}%
\providecommand \href [0]{\begingroup \@sanitize@url \@href}%
\providecommand \@href[1]{\@@startlink{#1}\@@href}%
\providecommand \@@href[1]{\endgroup#1\@@endlink}%
\providecommand \@sanitize@url [0]{\catcode `\\12\catcode `\$12\catcode
  `\&12\catcode `\#12\catcode `\^12\catcode `\_12\catcode `\%12\relax}%
\providecommand \@@startlink[1]{}%
\providecommand \@@endlink[0]{}%
\providecommand \url  [0]{\begingroup\@sanitize@url \@url }%
\providecommand \@url [1]{\endgroup\@href {#1}{\urlprefix }}%
\providecommand \urlprefix  [0]{URL }%
\providecommand \Eprint [0]{\href }%
\providecommand \doibase [0]{https://doi.org/}%
\providecommand \selectlanguage [0]{\@gobble}%
\providecommand \bibinfo  [0]{\@secondoftwo}%
\providecommand \bibfield  [0]{\@secondoftwo}%
\providecommand \translation [1]{[#1]}%
\providecommand \BibitemOpen [0]{}%
\providecommand \bibitemStop [0]{}%
\providecommand \bibitemNoStop [0]{.\EOS\space}%
\providecommand \EOS [0]{\spacefactor3000\relax}%
\providecommand \BibitemShut  [1]{\csname bibitem#1\endcsname}%
\let\auto@bib@innerbib\@empty
\bibitem [{\citenamefont {Nielsen}(1932)}]{nielsen32}%
  \BibitemOpen
  \bibfield  {author} {\bibinfo {author} {\bibfnamefont {H.~H.}\ \bibnamefont
  {Nielsen}},\ }\bibfield  {title} {\bibinfo {title} {The torsion
  oscillator-rotator in the quantum mechanics},\ }\href
  {https://doi.org/10.1103/PhysRev.40.445} {\bibfield  {journal} {\bibinfo
  {journal} {Phys.\ Rev.}\ }\textbf {\bibinfo {volume} {40}},\ \bibinfo {pages}
  {445} (\bibinfo {year} {1932})}\BibitemShut {NoStop}%
\bibitem [{\citenamefont {Gebbie}\ \emph {et~al.}(1963)\citenamefont {Gebbie},
  \citenamefont {Topping}, \citenamefont {Illsley},\ and\ \citenamefont
  {Dennison}}]{gebbie63}%
  \BibitemOpen
  \bibfield  {author} {\bibinfo {author} {\bibfnamefont {H.~A.}\ \bibnamefont
  {Gebbie}}, \bibinfo {author} {\bibfnamefont {G.}~\bibnamefont {Topping}},
  \bibinfo {author} {\bibfnamefont {R.}~\bibnamefont {Illsley}},\ and\ \bibinfo
  {author} {\bibfnamefont {D.~M.}\ \bibnamefont {Dennison}},\ }\bibfield
  {title} {\bibinfo {title} {The rotation spectrum of methyl alcohol from
  20~cm$^{-1}$ to 80~cm$^{-1}$},\ }\href
  {https://doi.org/10.1016/0022-2852(63)90019-5} {\bibfield  {journal}
  {\bibinfo  {journal} {J.\ Mol.\ Spectrosc.}\ }\textbf {\bibinfo {volume}
  {11}},\ \bibinfo {pages} {229} (\bibinfo {year} {1963})}\BibitemShut
  {NoStop}%
\bibitem [{\citenamefont {Lees}\ and\ \citenamefont {Baker}(1968)}]{lees68}%
  \BibitemOpen
  \bibfield  {author} {\bibinfo {author} {\bibfnamefont {R.~M.}\ \bibnamefont
  {Lees}}\ and\ \bibinfo {author} {\bibfnamefont {J.~G.}\ \bibnamefont
  {Baker}},\ }\bibfield  {title} {\bibinfo {title}
  {Torsion-{Vibration}–{Rotation} {Interactions} in {Methanol}. {I}.
  {Millimeter} {Wave} {Spectrum}},\ }\href {https://doi.org/10.1063/1.1668221}
  {\bibfield  {journal} {\bibinfo  {journal} {J.\ Chem.\ Phys.}\ }\textbf
  {\bibinfo {volume} {48}},\ \bibinfo {pages} {5299} (\bibinfo {year}
  {1968})}\BibitemShut {NoStop}%
\bibitem [{sup()}]{supplement}%
  \BibitemOpen
  \href@noop {} {}\bibinfo {note} {See Supplemental Material at [URL will be
  inserted by publisher], which includes Refs.~\cite{lin59, bethe57, curl65,
  wilson_decius_cross, xu08, A79}, for a figure showing the schematic energy
  level diagram of a non-rigid open-shell $^2\Sigma$ molecule displaying an
  internal rotation of its methyl group, for further details about the
  derivation of the quantum-mechanical Hamiltonian in Eq.~\eqref{hamilt}, for
  the alternate form of the RAM Hamiltonian in Eq.~\eqref{hrt}, and for the
  calculation of the electron spin-torsion energy levels shown in
  Fig.~\ref{psi_A_plot}.}\BibitemShut {Stop}%
\bibitem [{\citenamefont {Lin}\ and\ \citenamefont {Swalen}(1959)}]{lin59}%
  \BibitemOpen
  \bibfield  {author} {\bibinfo {author} {\bibfnamefont {C.~C.}\ \bibnamefont
  {Lin}}\ and\ \bibinfo {author} {\bibfnamefont {J.~D.}\ \bibnamefont
  {Swalen}},\ }\bibfield  {title} {\bibinfo {title} {Internal rotation and
  microwave spectroscopy},\ }\href {https://doi.org/10.1103/RevModPhys.31.841}
  {\bibfield  {journal} {\bibinfo  {journal} {Rev.\ Mod.\ Phys.}\ }\textbf
  {\bibinfo {volume} {31}},\ \bibinfo {pages} {841} (\bibinfo {year}
  {1959})}\BibitemShut {NoStop}%
\bibitem [{\citenamefont {Bethe}\ and\ \citenamefont
  {Salpeter}(1957)}]{bethe57}%
  \BibitemOpen
  \bibfield  {author} {\bibinfo {author} {\bibfnamefont {H.~A.}\ \bibnamefont
  {Bethe}}\ and\ \bibinfo {author} {\bibfnamefont {E.~E.}\ \bibnamefont
  {Salpeter}},\ }\href@noop {} {\emph {\bibinfo {title} {Handbuch der
  Physik}}},\ Vol.\ \bibinfo {volume} {XXXV}\ (\bibinfo  {publisher}
  {Springer},\ \bibinfo {address} {Berlin},\ \bibinfo {year} {1957})\ p.\
  \bibinfo {pages} {267},\ \bibinfo {note} {{E}q.~(39.14)}\BibitemShut
  {NoStop}%
\bibitem [{\citenamefont {{Curl, Jr.}}(1965)}]{curl65}%
  \BibitemOpen
  \bibfield  {author} {\bibinfo {author} {\bibfnamefont {R.~F.}\ \bibnamefont
  {{Curl, Jr.}}},\ }\bibfield  {title} {\bibinfo {title} {The relationship
  between electron spin rotation coupling constants and g-tensor components},\
  }\href {https://doi.org/10.1080/00268976500100761} {\bibfield  {journal}
  {\bibinfo  {journal} {Mol.\ Phys.}\ }\textbf {\bibinfo {volume} {9}},\
  \bibinfo {pages} {585} (\bibinfo {year} {1965})}\BibitemShut {NoStop}%
\bibitem [{\citenamefont {{Wilson, Jr.}}\ \emph {et~al.}(1955)\citenamefont
  {{Wilson, Jr.}}, \citenamefont {Decius},\ and\ \citenamefont
  {Cross}}]{wilson_decius_cross}%
  \BibitemOpen
  \bibfield  {author} {\bibinfo {author} {\bibfnamefont {E.~B.}\ \bibnamefont
  {{Wilson, Jr.}}}, \bibinfo {author} {\bibfnamefont {J.~C.}\ \bibnamefont
  {Decius}},\ and\ \bibinfo {author} {\bibfnamefont {P.~C.}\ \bibnamefont
  {Cross}},\ }\href@noop {} {\emph {\bibinfo {title} {Molecular Vibrations}}}\
  (\bibinfo  {publisher} {Mc Graw-Hill, Inc},\ \bibinfo {address} {New York,
  Toronto, London},\ \bibinfo {year} {1955})\BibitemShut {NoStop}%
\bibitem [{\citenamefont {Xu}\ \emph {et~al.}(2008)\citenamefont {Xu},
  \citenamefont {Fisher}, \citenamefont {Lees}, \citenamefont {Shi},
  \citenamefont {Hougen}, \citenamefont {Pearson}, \citenamefont {Drouin},
  \citenamefont {Blake},\ and\ \citenamefont {Braakman}}]{xu08}%
  \BibitemOpen
  \bibfield  {author} {\bibinfo {author} {\bibfnamefont {L.-H.}\ \bibnamefont
  {Xu}}, \bibinfo {author} {\bibfnamefont {J.}~\bibnamefont {Fisher}}, \bibinfo
  {author} {\bibfnamefont {R.~M.}\ \bibnamefont {Lees}}, \bibinfo {author}
  {\bibfnamefont {H.~Y.}\ \bibnamefont {Shi}}, \bibinfo {author} {\bibfnamefont
  {J.~T.}\ \bibnamefont {Hougen}}, \bibinfo {author} {\bibfnamefont {J.~C.}\
  \bibnamefont {Pearson}}, \bibinfo {author} {\bibfnamefont {B.~J.}\
  \bibnamefont {Drouin}}, \bibinfo {author} {\bibfnamefont {G.~A.}\
  \bibnamefont {Blake}},\ and\ \bibinfo {author} {\bibfnamefont
  {R.}~\bibnamefont {Braakman}},\ }\bibfield  {title} {\bibinfo {title}
  {Torsion–rotation global analysis of the first three torsional states
  ($v_t=0,$ 1, 2) and terahertz database for methanol},\ }\href
  {https://doi.org/10.1016/j.jms.2008.03.017} {\bibfield  {journal} {\bibinfo
  {journal} {J.\ Mol.\ Spectrosc.}\ }\textbf {\bibinfo {volume} {251}},\
  \bibinfo {pages} {305} (\bibinfo {year} {2008})}\BibitemShut {NoStop}%
\bibitem [{\citenamefont {Coudert}\ \emph {et~al.}(2015)\citenamefont
  {Coudert}, \citenamefont {{Gutl\'e}}, \citenamefont {Huet}, \citenamefont
  {Grabow},\ and\ \citenamefont {Levshakov}}]{A79}%
  \BibitemOpen
  \bibfield  {author} {\bibinfo {author} {\bibfnamefont {L.~H.}\ \bibnamefont
  {Coudert}}, \bibinfo {author} {\bibfnamefont {C.}~\bibnamefont {{Gutl\'e}}},
  \bibinfo {author} {\bibfnamefont {T.~R.}\ \bibnamefont {Huet}}, \bibinfo
  {author} {\bibfnamefont {J.-U.}\ \bibnamefont {Grabow}},\ and\ \bibinfo
  {author} {\bibfnamefont {S.~A.}\ \bibnamefont {Levshakov}},\ }\bibfield
  {title} {\bibinfo {title} {Spin-torsion effects in the hyperfine structure of
  methanol},\ }\href {https://doi.org/10.1063/1.4926942} {\bibfield  {journal}
  {\bibinfo  {journal} {J.\ Chem.\ Phys.}\ }\textbf {\bibinfo {volume} {143}},\
  \bibinfo {pages} {044304} (\bibinfo {year} {2015})}\BibitemShut {NoStop}%
\bibitem [{\citenamefont {Hirota}\ \emph {et~al.}(2007)\citenamefont {Hirota},
  \citenamefont {Mizoguchi}, \citenamefont {Ohshima}, \citenamefont {Katoh},
  \citenamefont {Sumiyoshi},\ and\ \citenamefont {Endo}}]{hirota07}%
  \BibitemOpen
  \bibfield  {author} {\bibinfo {author} {\bibfnamefont {E.}~\bibnamefont
  {Hirota}}, \bibinfo {author} {\bibfnamefont {A.}~\bibnamefont {Mizoguchi}},
  \bibinfo {author} {\bibfnamefont {Y.}~\bibnamefont {Ohshima}}, \bibinfo
  {author} {\bibfnamefont {K.}~\bibnamefont {Katoh}}, \bibinfo {author}
  {\bibfnamefont {Y.}~\bibnamefont {Sumiyoshi}},\ and\ \bibinfo {author}
  {\bibfnamefont {Y.}~\bibnamefont {Endo}},\ }\bibfield  {title} {\bibinfo
  {title} {Interplay of methyl-group internal rotation and fine and hyperfine
  interaction in a free radical: Fourier transform microwave spectroscopy of
  the acetyl radical},\ }\href {https://doi.org/10.1080/00268970601142624}
  {\bibfield  {journal} {\bibinfo  {journal} {Mol.\ Phys.}\ }\textbf {\bibinfo
  {volume} {105}},\ \bibinfo {pages} {455} (\bibinfo {year}
  {2007})}\BibitemShut {NoStop}%
\bibitem [{\citenamefont {Chahbazian}\ \emph
  {et~al.}(2024{\natexlab{a}})\citenamefont {Chahbazian}, \citenamefont
  {Martin-Drumel},\ and\ \citenamefont {Pirali}}]{chahbazian24}%
  \BibitemOpen
  \bibfield  {author} {\bibinfo {author} {\bibfnamefont {R.}~\bibnamefont
  {Chahbazian}}, \bibinfo {author} {\bibfnamefont {M.-A.}\ \bibnamefont
  {Martin-Drumel}},\ and\ \bibinfo {author} {\bibfnamefont {O.}~\bibnamefont
  {Pirali}},\ }\bibfield  {title} {\bibinfo {title} {High-resolution
  spectroscopic investigation of the \ce{CH2CHO} radical in the sub-millimeter
  region},\ }\href {https://doi.org/10.1021/acs.jpca.3c06326} {\bibfield
  {journal} {\bibinfo  {journal} {J.\ Phys.\ Chem.~A}\ }\textbf {\bibinfo
  {volume} {128}},\ \bibinfo {pages} {370} (\bibinfo {year}
  {2024}{\natexlab{a}})}\BibitemShut {NoStop}%
\bibitem [{\citenamefont {Coudert}\ \emph {et~al.}(2022)\citenamefont
  {Coudert}, \citenamefont {Chitarra}, \citenamefont {Spaniol}, \citenamefont
  {Loison}, \citenamefont {Martin-Drumel},\ and\ \citenamefont
  {Pirali}}]{A104}%
  \BibitemOpen
  \bibfield  {author} {\bibinfo {author} {\bibfnamefont {L.~H.}\ \bibnamefont
  {Coudert}}, \bibinfo {author} {\bibfnamefont {O.}~\bibnamefont {Chitarra}},
  \bibinfo {author} {\bibfnamefont {J.-C.}\ \bibnamefont {Spaniol}}, \bibinfo
  {author} {\bibfnamefont {J.-C.}\ \bibnamefont {Loison}}, \bibinfo {author}
  {\bibfnamefont {M.-A.}\ \bibnamefont {Martin-Drumel}},\ and\ \bibinfo
  {author} {\bibfnamefont {O.}~\bibnamefont {Pirali}},\ }\bibfield  {title}
  {\bibinfo {title} {Tunneling motion and splitting in the \ce{CH2OH} radical:
  (sub-)millimeter wave spectrum analysis},\ }\href
  {https://doi.org/10.1063/5.0095242} {\bibfield  {journal} {\bibinfo
  {journal} {J.\ Chem.\ Phys.}\ }\textbf {\bibinfo {volume} {156}},\ \bibinfo
  {pages} {244301} (\bibinfo {year} {2022})}\BibitemShut {NoStop}%
\bibitem [{\citenamefont {Zou}\ \emph {et~al.}(2020)\citenamefont {Zou},
  \citenamefont {Motiyenko}, \citenamefont {{Margul\`es}},\ and\ \citenamefont
  {Alekseev}}]{zou2020}%
  \BibitemOpen
  \bibfield  {author} {\bibinfo {author} {\bibfnamefont {L.}~\bibnamefont
  {Zou}}, \bibinfo {author} {\bibfnamefont {R.~A.}\ \bibnamefont {Motiyenko}},
  \bibinfo {author} {\bibfnamefont {L.}~\bibnamefont {{Margul\`es}}},\ and\
  \bibinfo {author} {\bibfnamefont {E.~A.}\ \bibnamefont {Alekseev}},\
  }\bibfield  {title} {\bibinfo {title} {{Millimeter-wave emission spectrometer
  based on direct digital synthesis}},\ }\href
  {https://doi.org/10.1063/5.0004461} {\bibfield  {journal} {\bibinfo
  {journal} {Rev.\ Sci.\ Instrum.}\ }\textbf {\bibinfo {volume} {91}},\
  \bibinfo {pages} {063104} (\bibinfo {year} {2020})}\BibitemShut {NoStop}%
\bibitem [{\citenamefont {Gyawali}(2023)}]{gyawali2023}%
  \BibitemOpen
  \bibfield  {author} {\bibinfo {author} {\bibfnamefont {P.}~\bibnamefont
  {Gyawali}},\ }\emph {\bibinfo {title} {Terahertz spectroscopy of molecules
  and molecular complexes of atmospheric interest exhibiting large amplitude
  motions}},\ \href {https://theses.hal.science/tel-04556686} {\bibinfo {type}
  {{PhD} thesis}},\ \bibinfo  {school} {University of Lille} (\bibinfo {year}
  {2023})\BibitemShut {NoStop}%
\bibitem [{\citenamefont {Hecht}\ and\ \citenamefont
  {Dennison}(1957)}]{hecht57}%
  \BibitemOpen
  \bibfield  {author} {\bibinfo {author} {\bibfnamefont {K.~T.}\ \bibnamefont
  {Hecht}}\ and\ \bibinfo {author} {\bibfnamefont {D.~M.}\ \bibnamefont
  {Dennison}},\ }\bibfield  {title} {\bibinfo {title} {Vibration‐hindered
  rotation interactions in methyl alcohol. the {$J=0\rightarrow 1$}
  transition},\ }\href {https://doi.org/10.1063/1.1743263} {\bibfield
  {journal} {\bibinfo  {journal} {J.\ Chem.\ Phys.}\ }\textbf {\bibinfo
  {volume} {26}},\ \bibinfo {pages} {48} (\bibinfo {year} {1957})}\BibitemShut
  {NoStop}%
\bibitem [{\citenamefont {Kirtman}(1962)}]{kirtman62}%
  \BibitemOpen
  \bibfield  {author} {\bibinfo {author} {\bibfnamefont {B.}~\bibnamefont
  {Kirtman}},\ }\bibfield  {title} {\bibinfo {title} {Interactions between
  ordinary vibrations and hindered internal rotation. {I.\ Rotational}
  energies},\ }\href {https://doi.org/10.1063/1.1733049} {\bibfield  {journal}
  {\bibinfo  {journal} {J.\ Chem.\ Phys.}\ }\textbf {\bibinfo {volume} {37}},\
  \bibinfo {pages} {2516} (\bibinfo {year} {1962})}\BibitemShut {NoStop}%
\bibitem [{\citenamefont {Herbst}\ \emph {et~al.}(1984)\citenamefont {Herbst},
  \citenamefont {Messer}, \citenamefont {{De Lucia}},\ and\ \citenamefont
  {Helminger}}]{herbst84}%
  \BibitemOpen
  \bibfield  {author} {\bibinfo {author} {\bibfnamefont {E.}~\bibnamefont
  {Herbst}}, \bibinfo {author} {\bibfnamefont {J.~K.}\ \bibnamefont {Messer}},
  \bibinfo {author} {\bibfnamefont {F.~C.}\ \bibnamefont {{De Lucia}}},\ and\
  \bibinfo {author} {\bibfnamefont {P.}~\bibnamefont {Helminger}},\ }\bibfield
  {title} {\bibinfo {title} {A new analysis and additional measurements of the
  millimiter and submillimeter spectrum of methanol},\ }\href
  {https://doi.org/10.1016/0022-2852(84)90285-6} {\bibfield  {journal}
  {\bibinfo  {journal} {J.\ Mol.\ Spectrosc.}\ }\textbf {\bibinfo {volume}
  {108}},\ \bibinfo {pages} {42} (\bibinfo {year} {1984})}\BibitemShut
  {NoStop}%
\bibitem [{\citenamefont {Hougen}\ \emph {et~al.}(1994)\citenamefont {Hougen},
  \citenamefont {Kleiner},\ and\ \citenamefont {Godefroid}}]{hougen94}%
  \BibitemOpen
  \bibfield  {author} {\bibinfo {author} {\bibfnamefont {J.~T.}\ \bibnamefont
  {Hougen}}, \bibinfo {author} {\bibfnamefont {I.}~\bibnamefont {Kleiner}},\
  and\ \bibinfo {author} {\bibfnamefont {M.}~\bibnamefont {Godefroid}},\
  }\bibfield  {title} {\bibinfo {title} {Selection rules and intensity
  calculations for a {$C_s$} asymmetric top molecule containing a methyl group
  internal rotor},\ }\href {https://doi.org/10.1006/jmsp.1994.1047} {\bibfield
  {journal} {\bibinfo  {journal} {J.\ Mol.\ Spectrosc.}\ }\textbf {\bibinfo
  {volume} {163}},\ \bibinfo {pages} {559} (\bibinfo {year}
  {1994})}\BibitemShut {NoStop}%
\bibitem [{\citenamefont {Brown}\ and\ \citenamefont {Sears}(1979)}]{brown79}%
  \BibitemOpen
  \bibfield  {author} {\bibinfo {author} {\bibfnamefont {J.~M.}\ \bibnamefont
  {Brown}}\ and\ \bibinfo {author} {\bibfnamefont {T.~J.}\ \bibnamefont
  {Sears}},\ }\bibfield  {title} {\bibinfo {title} {A reduced form of the
  spin-rotation {Hamiltonian} for asymmetric-top molecules, with applications
  to \ce{HO2} and \ce{NH2}},\ }\href
  {https://doi.org/10.1016/0022-2852(79)90153-X} {\bibfield  {journal}
  {\bibinfo  {journal} {J.\ Mol.\ Spectrosc.}\ }\textbf {\bibinfo {volume}
  {75}},\ \bibinfo {pages} {111} (\bibinfo {year} {1979})}\BibitemShut
  {NoStop}%
\bibitem [{\citenamefont {Bonah}\ \emph {et~al.}(2022)\citenamefont {Bonah},
  \citenamefont {Zingsheim}, \citenamefont {{M\"uller}}, \citenamefont
  {Guillemin}, \citenamefont {Lewen},\ and\ \citenamefont
  {Schlemmer}}]{Bonah2022}%
  \BibitemOpen
  \bibfield  {author} {\bibinfo {author} {\bibfnamefont {L.}~\bibnamefont
  {Bonah}}, \bibinfo {author} {\bibfnamefont {O.}~\bibnamefont {Zingsheim}},
  \bibinfo {author} {\bibfnamefont {H.~S.~P.}\ \bibnamefont {{M\"uller}}},
  \bibinfo {author} {\bibfnamefont {J.-C.}\ \bibnamefont {Guillemin}}, \bibinfo
  {author} {\bibfnamefont {F.}~\bibnamefont {Lewen}},\ and\ \bibinfo {author}
  {\bibfnamefont {S.}~\bibnamefont {Schlemmer}},\ }\bibfield  {title} {\bibinfo
  {title} {{LLWP}-a new loomis-wood software at the example of acetone-13c1},\
  }\href {https://doi.org/10.1016/j.jms.2022.111674} {\bibfield  {journal}
  {\bibinfo  {journal} {J.\ Mol.\ Spectrosc.}\ }\textbf {\bibinfo {volume}
  {388}},\ \bibinfo {pages} {111674} (\bibinfo {year} {2022})}\BibitemShut
  {NoStop}%
\bibitem [{\citenamefont {Brown}\ \emph {et~al.}(1980)\citenamefont {Brown},
  \citenamefont {Sears},\ and\ \citenamefont {Watson}}]{brown80}%
  \BibitemOpen
  \bibfield  {author} {\bibinfo {author} {\bibfnamefont {J.~M.}\ \bibnamefont
  {Brown}}, \bibinfo {author} {\bibfnamefont {T.~J.}\ \bibnamefont {Sears}},\
  and\ \bibinfo {author} {\bibfnamefont {J.~K.~G.}\ \bibnamefont {Watson}},\
  }\bibfield  {title} {\bibinfo {title} {The isotopic dependence of the
  spin-rotation interaction for an asymmetric top molecule},\ }\href
  {https://doi.org/10.1080/00268978000102661} {\bibfield  {journal} {\bibinfo
  {journal} {Mol.\ Phys.}\ }\textbf {\bibinfo {volume} {41}},\ \bibinfo {pages}
  {173} (\bibinfo {year} {1980})}\BibitemShut {NoStop}%
\bibitem [{\citenamefont {Brown}\ and\ \citenamefont {Watson}(1977)}]{brown77}%
  \BibitemOpen
  \bibfield  {author} {\bibinfo {author} {\bibfnamefont {J.~M.}\ \bibnamefont
  {Brown}}\ and\ \bibinfo {author} {\bibfnamefont {J.~K.~G.}\ \bibnamefont
  {Watson}},\ }\bibfield  {title} {\bibinfo {title} {Spin-orbit and
  spin-rotation coupling in doublet states of diatomic molecules},\ }\href
  {https://doi.org/10.1016/0022-2852(77)90358-7} {\bibfield  {journal}
  {\bibinfo  {journal} {J.\ Mol.\ Spectrosc.}\ }\textbf {\bibinfo {volume}
  {65}},\ \bibinfo {pages} {65} (\bibinfo {year} {1977})}\BibitemShut {NoStop}%
\bibitem [{\citenamefont {Burkhard}\ and\ \citenamefont
  {Dennison}(1959)}]{burkhard59}%
  \BibitemOpen
  \bibfield  {author} {\bibinfo {author} {\bibfnamefont {D.~G.}\ \bibnamefont
  {Burkhard}}\ and\ \bibinfo {author} {\bibfnamefont {D.~M.}\ \bibnamefont
  {Dennison}},\ }\bibfield  {title} {\bibinfo {title} {Rotation spectrum of
  methyl alcohol},\ }\href {https://doi.org/10.1016/0022-2852(59)90030-X}
  {\bibfield  {journal} {\bibinfo  {journal} {J.\ Mol.\ Spectrosc.}\ }\textbf
  {\bibinfo {volume} {3}},\ \bibinfo {pages} {299} (\bibinfo {year}
  {1959})}\BibitemShut {NoStop}%
\bibitem [{\citenamefont {Ferrero}\ \emph {et~al.}(2023)\citenamefont
  {Ferrero}, \citenamefont {Ceccarelli}, \citenamefont {Ugliengo},
  \citenamefont {Sodupe},\ and\ \citenamefont {Rimola}}]{Ferrero2023}%
  \BibitemOpen
  \bibfield  {author} {\bibinfo {author} {\bibfnamefont {S.}~\bibnamefont
  {Ferrero}}, \bibinfo {author} {\bibfnamefont {C.}~\bibnamefont {Ceccarelli}},
  \bibinfo {author} {\bibfnamefont {P.}~\bibnamefont {Ugliengo}}, \bibinfo
  {author} {\bibfnamefont {M.}~\bibnamefont {Sodupe}},\ and\ \bibinfo {author}
  {\bibfnamefont {A.}~\bibnamefont {Rimola}},\ }\bibfield  {title} {\bibinfo
  {title} {Formation of complex organic molecules on interstellar {CO} ices?
  {Insights} from computational chemistry simulations},\ }\href
  {https://doi.org/10.3847/1538-4357/acd192} {\bibfield  {journal} {\bibinfo
  {journal} {Astrophys.\ J.}\ }\textbf {\bibinfo {volume} {951}},\ \bibinfo
  {pages} {150} (\bibinfo {year} {2023})}\BibitemShut {NoStop}%
\bibitem [{\citenamefont {Feldman}(2023)}]{ijms241914510}%
  \BibitemOpen
  \bibfield  {author} {\bibinfo {author} {\bibfnamefont {V.~I.}\ \bibnamefont
  {Feldman}},\ }\bibfield  {title} {\bibinfo {title} {Astrochemically relevant
  radicals and radical–molecule complexes: A new insight from matrix
  isolation},\ }\href {https://doi.org/10.3390/ijms241914510} {\bibfield
  {journal} {\bibinfo  {journal} {Int.\ J.\ Molec.\ Sciences}\ }\textbf
  {\bibinfo {volume} {24}},\ \bibinfo {pages} {14510} (\bibinfo {year}
  {2023})}\BibitemShut {NoStop}%
\bibitem [{\citenamefont {Cernicharo}\ \emph {et~al.}(2021)\citenamefont
  {Cernicharo}, \citenamefont {Cabezas}, \citenamefont {Bailleux},
  \citenamefont {{Margul\`es}}, \citenamefont {Motiyenko}, \citenamefont {Zou},
  \citenamefont {Endo}, \citenamefont {{Berm\'udez}}, \citenamefont
  {{Ag\'undez}}, \citenamefont {Marcelino}, \citenamefont {Lefloch},
  \citenamefont {Tercero},\ and\ \citenamefont {{de Vicente}}}]{cernicharo21}%
  \BibitemOpen
  \bibfield  {author} {\bibinfo {author} {\bibfnamefont {J.}~\bibnamefont
  {Cernicharo}}, \bibinfo {author} {\bibfnamefont {C.}~\bibnamefont {Cabezas}},
  \bibinfo {author} {\bibfnamefont {S.}~\bibnamefont {Bailleux}}, \bibinfo
  {author} {\bibfnamefont {L.}~\bibnamefont {{Margul\`es}}}, \bibinfo {author}
  {\bibfnamefont {R.}~\bibnamefont {Motiyenko}}, \bibinfo {author}
  {\bibfnamefont {L.}~\bibnamefont {Zou}}, \bibinfo {author} {\bibfnamefont
  {Y.}~\bibnamefont {Endo}}, \bibinfo {author} {\bibfnamefont {C.}~\bibnamefont
  {{Berm\'udez}}}, \bibinfo {author} {\bibfnamefont {M.}~\bibnamefont
  {{Ag\'undez}}}, \bibinfo {author} {\bibfnamefont {N.}~\bibnamefont
  {Marcelino}}, \bibinfo {author} {\bibfnamefont {B.}~\bibnamefont {Lefloch}},
  \bibinfo {author} {\bibfnamefont {B.}~\bibnamefont {Tercero}},\ and\ \bibinfo
  {author} {\bibfnamefont {P.}~\bibnamefont {{de Vicente}}},\ }\bibfield
  {title} {\bibinfo {title} {Discovery of the acetyl cation, \ce{CH3CO+}, in
  space and in the laboratory},\ }\href
  {https://doi.org/10.1051/0004-6361/202040076} {\bibfield  {journal} {\bibinfo
   {journal} {A\&A}\ }\textbf {\bibinfo {volume} {646}},\ \bibinfo {pages} {L7}
  (\bibinfo {year} {2021})}\BibitemShut {NoStop}%
\bibitem [{\citenamefont {Herbst}\ and\ \citenamefont {van
  Dishoeck}(2009)}]{herbst2009}%
  \BibitemOpen
  \bibfield  {author} {\bibinfo {author} {\bibfnamefont {E.}~\bibnamefont
  {Herbst}}\ and\ \bibinfo {author} {\bibfnamefont {E.~F.}\ \bibnamefont {van
  Dishoeck}},\ }\bibfield  {title} {\bibinfo {title} {Complex organic
  interstellar molecules},\ }\href
  {https://doi.org/https://doi.org/10.1146/annurev-astro-082708-101654}
  {\bibfield  {journal} {\bibinfo  {journal} {Ann.\ Rev.\ Astron.\ Astrophys.}\
  }\textbf {\bibinfo {volume} {47}},\ \bibinfo {pages} {427} (\bibinfo {year}
  {2009})}\BibitemShut {NoStop}%
\bibitem [{\citenamefont {Butscher}\ \emph {et~al.}(2017)\citenamefont
  {Butscher}, \citenamefont {Duvernay}, \citenamefont {Rimola}, \citenamefont
  {Segado-Centellas},\ and\ \citenamefont {Chiavassa}}]{C6CP07024H}%
  \BibitemOpen
  \bibfield  {author} {\bibinfo {author} {\bibfnamefont {T.}~\bibnamefont
  {Butscher}}, \bibinfo {author} {\bibfnamefont {F.}~\bibnamefont {Duvernay}},
  \bibinfo {author} {\bibfnamefont {A.}~\bibnamefont {Rimola}}, \bibinfo
  {author} {\bibfnamefont {M.}~\bibnamefont {Segado-Centellas}},\ and\ \bibinfo
  {author} {\bibfnamefont {T.}~\bibnamefont {Chiavassa}},\ }\bibfield  {title}
  {\bibinfo {title} {Radical recombination in interstellar ices, a not so
  simple mechanism},\ }\href {https://doi.org/10.1039/C6CP07024H} {\bibfield
  {journal} {\bibinfo  {journal} {Phys.\ Chem.\ Chem.\ Phys.}\ }\textbf
  {\bibinfo {volume} {19}},\ \bibinfo {pages} {2857} (\bibinfo {year}
  {2017})}\BibitemShut {NoStop}%
\bibitem [{\citenamefont {Chahbazian}\ \emph
  {et~al.}(2024{\natexlab{b}})\citenamefont {Chahbazian}, \citenamefont
  {Juppet},\ and\ \citenamefont {Pirali}}]{Chahbazian2024}%
  \BibitemOpen
  \bibfield  {author} {\bibinfo {author} {\bibfnamefont {R.}~\bibnamefont
  {Chahbazian}}, \bibinfo {author} {\bibfnamefont {L.}~\bibnamefont {Juppet}},\
  and\ \bibinfo {author} {\bibfnamefont {O.}~\bibnamefont {Pirali}},\
  }\bibfield  {title} {\bibinfo {title} {Unveiling the spectroscopy of complex
  organic radicals by exploiting faraday rotation at (sub-)millimeter
  wavelengths. {Illustration} with the acetonyl radical},\ }\href
  {https://doi.org/10.1021/acs.jpclett.4c01936} {\bibfield  {journal} {\bibinfo
   {journal} {The J.\ Phys.\ Chem.\ Lett.}\ }\textbf {\bibinfo {volume} {15}},\
  \bibinfo {pages} {9803} (\bibinfo {year} {2024}{\natexlab{b}})}\BibitemShut
  {NoStop}%
\bibitem [{\citenamefont {Kasai}\ \emph {et~al.}(2011)\citenamefont {Kasai},
  \citenamefont {Dupuy}, \citenamefont {Saito}, \citenamefont {Hashimoto},
  \citenamefont {Sabu}, \citenamefont {Kondo}, \citenamefont {Sumiyoshi},\ and\
  \citenamefont {Endo}}]{acp-11-8607-2011}%
  \BibitemOpen
  \bibfield  {author} {\bibinfo {author} {\bibfnamefont {Y.}~\bibnamefont
  {Kasai}}, \bibinfo {author} {\bibfnamefont {E.}~\bibnamefont {Dupuy}},
  \bibinfo {author} {\bibfnamefont {R.}~\bibnamefont {Saito}}, \bibinfo
  {author} {\bibfnamefont {K.}~\bibnamefont {Hashimoto}}, \bibinfo {author}
  {\bibfnamefont {A.}~\bibnamefont {Sabu}}, \bibinfo {author} {\bibfnamefont
  {S.}~\bibnamefont {Kondo}}, \bibinfo {author} {\bibfnamefont
  {Y.}~\bibnamefont {Sumiyoshi}},\ and\ \bibinfo {author} {\bibfnamefont
  {Y.}~\bibnamefont {Endo}},\ }\bibfield  {title} {\bibinfo {title} {The
  \ce{H2O-O2} water vapour complex in the earth's atmosphere},\ }\href
  {https://doi.org/10.5194/acp-11-8607-2011} {\bibfield  {journal} {\bibinfo
  {journal} {Atmos.\ Chem.\ Phys.}\ }\textbf {\bibinfo {volume} {11}},\
  \bibinfo {pages} {8607} (\bibinfo {year} {2011})}\BibitemShut {NoStop}%
\end{thebibliography}%


\begin{thebibliography}{6}%
\makeatletter
\providecommand \@ifxundefined [1]{%
 \@ifx{#1\undefined}
}%
\providecommand \@ifnum [1]{%
 \ifnum #1\expandafter \@firstoftwo
 \else \expandafter \@secondoftwo
 \fi
}%
\providecommand \@ifx [1]{%
 \ifx #1\expandafter \@firstoftwo
 \else \expandafter \@secondoftwo
 \fi
}%
\providecommand \natexlab [1]{#1}%
\providecommand \enquote  [1]{``#1''}%
\providecommand \bibnamefont  [1]{#1}%
\providecommand \bibfnamefont [1]{#1}%
\providecommand \citenamefont [1]{#1}%
\providecommand \href@noop [0]{\@secondoftwo}%
\providecommand \href [0]{\begingroup \@sanitize@url \@href}%
\providecommand \@href[1]{\@@startlink{#1}\@@href}%
\providecommand \@@href[1]{\endgroup#1\@@endlink}%
\providecommand \@sanitize@url [0]{\catcode `\\12\catcode `\$12\catcode
  `\&12\catcode `\#12\catcode `\^12\catcode `\_12\catcode `\%12\relax}%
\providecommand \@@startlink[1]{}%
\providecommand \@@endlink[0]{}%
\providecommand \url  [0]{\begingroup\@sanitize@url \@url }%
\providecommand \@url [1]{\endgroup\@href {#1}{\urlprefix }}%
\providecommand \urlprefix  [0]{URL }%
\providecommand \Eprint [0]{\href }%
\providecommand \doibase [0]{https://doi.org/}%
\providecommand \selectlanguage [0]{\@gobble}%
\providecommand \bibinfo  [0]{\@secondoftwo}%
\providecommand \bibfield  [0]{\@secondoftwo}%
\providecommand \translation [1]{[#1]}%
\providecommand \BibitemOpen [0]{}%
\providecommand \bibitemStop [0]{}%
\providecommand \bibitemNoStop [0]{.\EOS\space}%
\providecommand \EOS [0]{\spacefactor3000\relax}%
\providecommand \BibitemShut  [1]{\csname bibitem#1\endcsname}%
\let\auto@bib@innerbib\@empty
\bibitem [{\citenamefont {Lin}\ and\ \citenamefont {Swalen}(1959)}]{lin59}%
  \BibitemOpen
  \bibfield  {author} {\bibinfo {author} {\bibfnamefont {C.~C.}\ \bibnamefont
  {Lin}}\ and\ \bibinfo {author} {\bibfnamefont {J.~D.}\ \bibnamefont
  {Swalen}},\ }\bibfield  {title} {\bibinfo {title} {Internal rotation and
  microwave spectroscopy},\ }\href {https://doi.org/10.1103/RevModPhys.31.841}
  {\bibfield  {journal} {\bibinfo  {journal} {Rev.\ Mod.\ Phys.}\ }\textbf
  {\bibinfo {volume} {31}},\ \bibinfo {pages} {841} (\bibinfo {year}
  {1959})}\BibitemShut {NoStop}%
\bibitem [{\citenamefont {Bethe}\ and\ \citenamefont
  {Salpeter}(1957)}]{bethe57}%
  \BibitemOpen
  \bibfield  {author} {\bibinfo {author} {\bibfnamefont {H.~A.}\ \bibnamefont
  {Bethe}}\ and\ \bibinfo {author} {\bibfnamefont {E.~E.}\ \bibnamefont
  {Salpeter}},\ }\href@noop {} {\emph {\bibinfo {title} {Handbuch der
  Physik}}},\ Vol.\ \bibinfo {volume} {XXXV}\ (\bibinfo  {publisher}
  {Springer},\ \bibinfo {address} {Berlin},\ \bibinfo {year} {1957})\ p.\
  \bibinfo {pages} {267},\ \bibinfo {note} {{E}q.~(39.14)}\BibitemShut
  {NoStop}%
\bibitem [{\citenamefont {{Curl, Jr.}}(1965)}]{curl65}%
  \BibitemOpen
  \bibfield  {author} {\bibinfo {author} {\bibfnamefont {R.~F.}\ \bibnamefont
  {{Curl, Jr.}}},\ }\bibfield  {title} {\bibinfo {title} {The relationship
  between electron spin rotation coupling constants and g-tensor components},\
  }\href {https://doi.org/10.1080/00268976500100761} {\bibfield  {journal}
  {\bibinfo  {journal} {Mol.\ Phys.}\ }\textbf {\bibinfo {volume} {9}},\
  \bibinfo {pages} {585} (\bibinfo {year} {1965})}\BibitemShut {NoStop}%
\bibitem [{\citenamefont {{Wilson, Jr.}}\ \emph {et~al.}(1955)\citenamefont
  {{Wilson, Jr.}}, \citenamefont {Decius},\ and\ \citenamefont
  {Cross}}]{wilson_decius_cross}%
  \BibitemOpen
  \bibfield  {author} {\bibinfo {author} {\bibfnamefont {E.~B.}\ \bibnamefont
  {{Wilson, Jr.}}}, \bibinfo {author} {\bibfnamefont {J.~C.}\ \bibnamefont
  {Decius}},\ and\ \bibinfo {author} {\bibfnamefont {P.~C.}\ \bibnamefont
  {Cross}},\ }\href@noop {} {\emph {\bibinfo {title} {Molecular Vibrations}}}\
  (\bibinfo  {publisher} {Mc Graw-Hill, Inc},\ \bibinfo {address} {New York,
  Toronto, London},\ \bibinfo {year} {1955})\BibitemShut {NoStop}%
\bibitem [{\citenamefont {Xu}\ \emph {et~al.}(2008)\citenamefont {Xu},
  \citenamefont {Fisher}, \citenamefont {Lees}, \citenamefont {Shi},
  \citenamefont {Hougen}, \citenamefont {Pearson}, \citenamefont {Drouin},
  \citenamefont {Blake},\ and\ \citenamefont {Braakman}}]{xu08}%
  \BibitemOpen
  \bibfield  {author} {\bibinfo {author} {\bibfnamefont {L.-H.}\ \bibnamefont
  {Xu}}, \bibinfo {author} {\bibfnamefont {J.}~\bibnamefont {Fisher}}, \bibinfo
  {author} {\bibfnamefont {R.~M.}\ \bibnamefont {Lees}}, \bibinfo {author}
  {\bibfnamefont {H.~Y.}\ \bibnamefont {Shi}}, \bibinfo {author} {\bibfnamefont
  {J.~T.}\ \bibnamefont {Hougen}}, \bibinfo {author} {\bibfnamefont {J.~C.}\
  \bibnamefont {Pearson}}, \bibinfo {author} {\bibfnamefont {B.~J.}\
  \bibnamefont {Drouin}}, \bibinfo {author} {\bibfnamefont {G.~A.}\
  \bibnamefont {Blake}},\ and\ \bibinfo {author} {\bibfnamefont
  {R.}~\bibnamefont {Braakman}},\ }\bibfield  {title} {\bibinfo {title}
  {Torsion–rotation global analysis of the first three torsional states
  ($v_t=0,$ 1, 2) and terahertz database for methanol},\ }\href
  {https://doi.org/10.1016/j.jms.2008.03.017} {\bibfield  {journal} {\bibinfo
  {journal} {J.\ Mol.\ Spectrosc.}\ }\textbf {\bibinfo {volume} {251}},\
  \bibinfo {pages} {305} (\bibinfo {year} {2008})}\BibitemShut {NoStop}%
\bibitem [{\citenamefont {Coudert}\ \emph {et~al.}(2015)\citenamefont
  {Coudert}, \citenamefont {{Gutl\'e}}, \citenamefont {Huet}, \citenamefont
  {Grabow},\ and\ \citenamefont {Levshakov}}]{A79}%
  \BibitemOpen
  \bibfield  {author} {\bibinfo {author} {\bibfnamefont {L.~H.}\ \bibnamefont
  {Coudert}}, \bibinfo {author} {\bibfnamefont {C.}~\bibnamefont {{Gutl\'e}}},
  \bibinfo {author} {\bibfnamefont {T.~R.}\ \bibnamefont {Huet}}, \bibinfo
  {author} {\bibfnamefont {J.-U.}\ \bibnamefont {Grabow}},\ and\ \bibinfo
  {author} {\bibfnamefont {S.~A.}\ \bibnamefont {Levshakov}},\ }\bibfield
  {title} {\bibinfo {title} {Spin-torsion effects in the hyperfine structure of
  methanol},\ }\href {https://doi.org/10.1063/1.4926942} {\bibfield  {journal}
  {\bibinfo  {journal} {J.\ Chem.\ Phys.}\ }\textbf {\bibinfo {volume} {143}},\
  \bibinfo {pages} {044304} (\bibinfo {year} {2015})}\BibitemShut {NoStop}%
\end{thebibliography}%

\end{document}


\title[Electron spin-torsion coupling]{{\Large Supplemental Material}\vspace{3.0ex}\\
{\rm Evidence for electron spin-torsion coupling\\in
the rotational spectrum of the \ce{CH3CO} radical}}
\author{L. H. Coudert, O. Pirali, M.-A. Martin-Drumel, R. Chahbazian, L. Zou, R. A. Motiyenko, and L. Margul\`es}

\maketitle\setcounter{subsection}{1}%

\subsection{\thesubsection~Schematic energy level diagram}
\begin{figure*}
\includegraphics[width=0.6\linewidth]{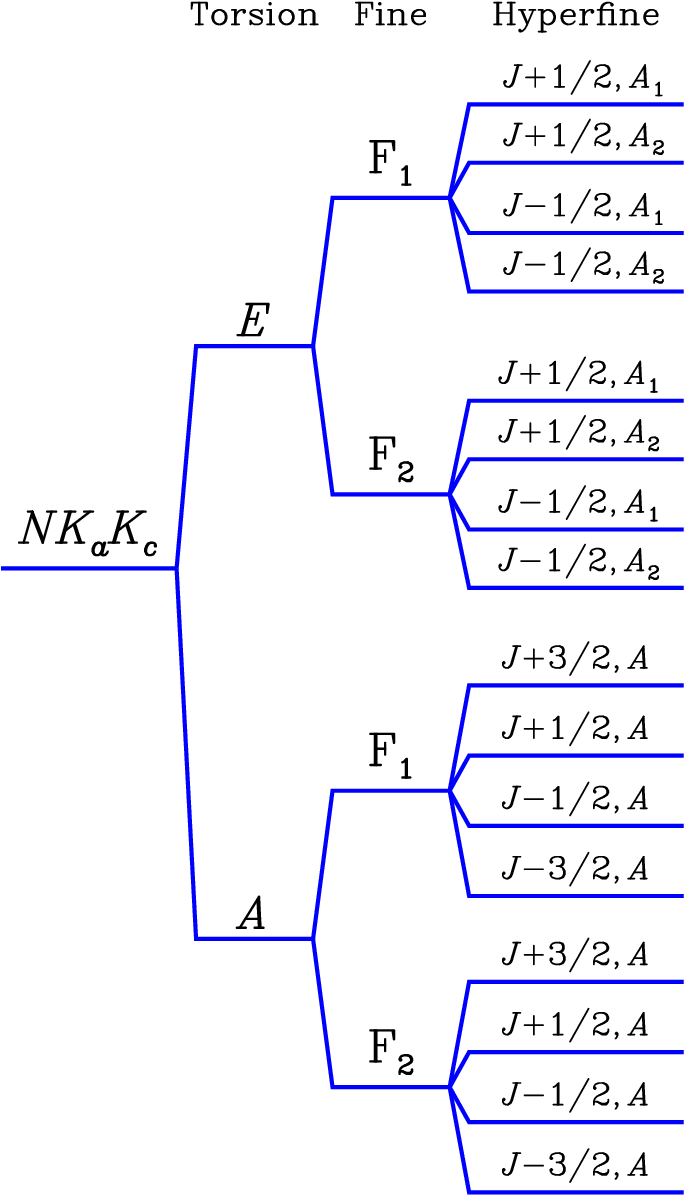}
\caption{\label{params}Schematic energy level diagram of a
non-rigid $^2\Sigma$ open shell molecule displaying an internal rotation
of its methyl group. The cumulated effects of the torsion
of the methyl group, of the fine interaction, and of the
hyperfine coupling are taken into account.  Starting from a
rotational level labeled $NK_aK_c$, the torsional motion is
taken into account leading to a splitting into 2 sub-levels
labeled with their $C_{3v}$ symmetry species.  $A$ is either
$A_1$ or $A_2$ depending on $NK_aK_c$.  The fine interaction
leads to a splitting of the rotation-torsion energy levels into
F$_1$ and F$_2$ sub-levels depending on whether $J$ is $N+1/2$
of $N-1/2$, respectively. These levels are further split into
several sub-levels by the hyperfine coupling. It is assumed
here that this coupling arises only from the three hydrogen
atoms of the methyl group.  Hyperfine levels are labeled with
the hyperfine quantum number $F$ corresponding to the coupling
scheme ${\bf F} = {\bf J} + {\bf I}$, where ${\bf I}$ is the
total nuclear spin.  For doubly degenerate $E$-type levels,
hyperfine levels levels are characterized by $F=J\pm1/2$
and belong either to the $A_1$ or the $A_2$ symmetry species
of $C_{3v}$~\cite{A79}. For nondegenerate $A$-type levels,
$F=J\pm3/2$ and $J\pm1/2$.}\end{figure*}

\stepcounter{subsection}%
The schematic energy level diagram of a non-rigid $^2\Sigma$
open shell molecule displaying an internal rotation of its
methyl group, like the acetyl radical, can be found in Fig.~\ref{params}.  This figure
shows the cumulated effects of the torsional motion, of the
fine interaction, and of the hyperfine coupling.

\subsection{\thesubsection~Derivation of the classical Lagrangian}\stepcounter{subsection}

Equation~(1) of the \pap\ is written in terms of the molecule
fixed Cartesian coordinates of the $n_e$ electrons, of the
angle of internal rotation $\gamma$, and of the molecule fixed
components of the angular velocity vector $\boldsymbol{\omega}$.
The second line of Eq.~(1) of the \pap, which gives the kinetic
energy for the $n_n$ nuclei, is a generalized expression of
Eq.~(2-36) of Lin and Swalen~\cite{lin59} which is restricted
to the case when the inertia tensor is diagonal.  From Eq.~(1)
of the \pap, the Lagrangian $\mathcal{L}$ in Eq.~(2) of the
\pap\ is obtained by adding magnetic effects as well as the
terms from Beth and Salpeter~\cite{bethe57}.  Magnetic effects
are described by the usual term $q_i {\bf A}_i \cdot {\bf
v}_i$, where ${\bf A}_i$ is the potential vector at particle
$i$ with charge $q_i$ and speed ${\bf v}_i$.  In the molecule
fixed axis system, the speed needs to be changed into:
\begin{equation}
\dot{\bf r}_i + \boldsymbol{\omega} \times {\bf r}_i
\label{speed}
\end{equation}
The first term in this equation gives rise to the following
terms in the Lagrangian $\mathcal{L}$ in Eq.~(2) of the
\pap:
\begin{equation}
\dot{\gamma} S_{\gamma}\mbox{ \ and \ }
\sum_i (-e/c) {\bf A}_i\cdot \dot{\bf r}_i
\label{fterms}
\end{equation}
where the first term is for the nuclei and $S_{\gamma}$
is defined in the second of Eqs.~(3) of the \pap.
The second term in Eq.~\eqref{speed} leads in
the Lagrangian $\mathcal{L}$ to the terms:
\begin{equation}
\omega \cdot {\bf S}_{\rm R} \mbox{ \ and \ }
\omega \cdot \sum_i(-e/c) {\bf r}_i \times {\bf A}_i
\end{equation}
where the first term is for the nuclei
and ${\bf S}_{\rm R}$ is given in the first of Eqs.~(3) of the
\pap. The three remaining terms in Eq.~(2) of the \pap\
involve the electric field  $\boldsymbol{\mathcal{E}}_i$
and come from Beth and Salpeter~\cite{bethe57}.

\subsection{\thesubsection~Derivation of the classical Hamiltonian}\stepcounter{subsection}

The Hamiltonian in Eq.~(4) of the \pap\
is retrieved from the Lagrangian using
Eq.~(13) of Curl~\cite{curl65} which is a fundamental physic
equation. In the present case, we have three sets of velocities:
$\dot{\gamma}$, $\dot{\bf r}_i$, and $\boldsymbol{\omega}$
and their conjugate momenta are
determined with the usual procedure.
We first obtain the rotational angular momentum ${\bf N}$
conjugate to the angular velocity $\boldsymbol{\omega}$ by:
\begin{equation}
\begin{split}
{\bf N} & = \frac{\partial\mathcal{L}}{\partial\boldsymbol{\omega}}=
{\bf I}\cdot\boldsymbol{\omega} + \sum_i m ({\bf r}_i \times \dot{\bf r}_i) + \dot{\gamma} {\bf C}_{\gamma} + {\bf S}_{\rm R}\vspace{1.0ex}\\
&-\sum_i {\bf r}_i \times {\bf B}_i
\end{split}
\label{rotangular}
\end{equation}
where ${\bf I} = {\bf I}_n + {\bf I}_e$ and
${\bf B}_i$ is the vector $(e/c) {\bf A}_i + (\mu_{\rm B}/c) ({\bf S}_i \times \boldsymbol{\mathcal{E}}_i)$.
${\bf p}_i$, the  momentum conjugate to $\dot{\bf r}_i$ the speed of electron $i$, is obtained by:
\begin{equation}
{\bf p}_i = \frac{\partial\mathcal{L}}{\partial\dot{\bf r}_i}=
m \dot{\bf r}_i + m \boldsymbol{\omega} \times {\bf r}_i - {\bf B}_i
\label{ppi}
\end{equation}
The internal rotation angular momentum $p_{\gamma}$, conjugate
to $\dot{\gamma}$, is obtained by:
\begin{equation}
{\bf p}_{\gamma} = \dot{\gamma} \ineta + \boldsymbol{\omega} \cdot {\bf C}_{\gamma} + S_{\gamma}
\label{gammad}
\end{equation}
Inserting into Eq.~\eqref{rotangular} the expression of $\dot{\bf r}_i$ from Eq.~\eqref{ppi} and
that of $\dot{\gamma}$ from Eq.~\eqref{gammad}, we obtain:
\begin{equation}
{\bf N} = {\bf I}_n \cdot \boldsymbol{\omega} + {\bf L} + {\bf S}_{\rm R}
+ ({\bf C}_{\gamma} p_{\gamma}
- {\bf C}_{\gamma} {\bf C}_{\gamma}^{\intercal}  \cdot \boldsymbol{\omega}
- {\bf C}_{\gamma} S_{\gamma})/\ineta
\end{equation}
where ${\bf L} = \sum_i m ({\bf r}_i \times {\bf p}_i)$. From
this equation, $\boldsymbol{\omega}$ can be expressed as:
\begin{equation}
\boldsymbol{\omega} = \boldsymbol{\mu} \cdot ({\bf N} - {\bf L} - p_{\gamma} \boldsymbol{\lambda} + S_{\gamma} \boldsymbol{\lambda} - {\bf S}_{\rm R})
\label{omegadef}
\end{equation}
where $\boldsymbol{\mu}$ is the $3\times 3$ tensor introduced after
Eq.~(4) of the \pap\ equal to the inverse of
${\bf I}_n -{\bf C}_{\gamma} {\bf C}_{\gamma}^{\intercal} / \ineta =
{\bf I}_n - \boldsymbol{\lambda} \boldsymbol{\lambda}^{\intercal} \ineta$.

In agreement with Eq.~(13) of Curl~\cite{curl65}, we evaluate first:
\begin{equation}
E=\boldsymbol{\omega}\cdot {\bf N} + \sum_i {\bf p}_i \cdot \dot{\bf r}_i + p_{\gamma} \dot{\gamma}
\label{edef}
\end{equation}
Equations~\eqref{ppi}, \eqref{gammad}, and \eqref{omegadef} are used to calculate $E$
which is expressed using mostly conjugate momenta:
\begin{equation}
\begin{split}
E & = p_{\gamma} (p_{\gamma} - S_{\gamma}) / \ineta  -  p_{\gamma} \boldsymbol{\lambda} \cdot \boldsymbol{\omega} + {\bf N}\cdot \boldsymbol{\omega}\vspace{1.0ex}\\
&+ \sum_i {\bf p}_i\cdot ({\bf p}_i + {\bf B}_i)/m - {\bf L}\cdot \boldsymbol{\omega}
\end{split}
\end{equation}
Rearranging the terms in this equation leads to:
\begin{equation}
\begin{split}
E&= ({\bf N} - {\bf L} - p_{\gamma} \boldsymbol{\lambda} + S_{\gamma}  \boldsymbol{\lambda} - {\bf S}_{\rm R})\cdot \boldsymbol{\omega}\vspace{1.0ex}\\
& + \boldsymbol{\omega} \cdot ({\bf S}_{\rm R} - S_{\gamma} \boldsymbol{\lambda}) + (p_{\gamma} - S_{\gamma})^2/ \ineta \vspace{1.0ex}\\
& + S_{\gamma} (p_{\gamma} - S_{\gamma})/\ineta + \sum_i ({\bf p}_i + {\bf B}_i)^2 /m\vspace{1.0ex}\\
&- \sum_i ({\bf p}_i\cdot {\bf B}_i + {\bf B}_i^2) /m\vspace{1.0ex}\\
&= 2H - 2 V + \sum_i (\mu_{\rm B}^2/mc^2) ({\bf S}_i \times \boldsymbol{\mathcal{E}}_i)^2\vspace{1.0ex}\\
& + \boldsymbol{\omega} \cdot S_{\rm R} + S_{\gamma} \dot{\gamma}
-\sum_i {\bf B}_i\cdot (m \dot{\bf r}_i + \boldsymbol{\omega} \times {\bf r}_i)
\end{split}
\label{eham}
\end{equation}
where $H$ is the Hamiltonian in Eq.~(4) of the \pap\ and $V$
is the potential energy function as defined for Eq.~(2) of
the \pap.  When expressing $E$ in Eq.~\eqref{edef} without conjugate momenta,
the following expression arises:
\begin{equation}
\begin{split}
E & = \boldsymbol{\omega} \cdot {\bf I} \cdot \boldsymbol{\omega} + \boldsymbol{\omega} \cdot \sum_i m ({\bf r}_i \times \dot{\bf r}_i)
+ \dot{\gamma}  \boldsymbol{\omega} \cdot {\bf C}_{\gamma}\vspace{1.0ex}\\
& + \boldsymbol{\omega} \cdot S_{\rm R} -  \sum_i ({\bf r}_i \times {\bf B}_i) \cdot \boldsymbol{\omega} \vspace{1.0ex}\\
& + \sum_i \dot{\bf r}_i \cdot (m \dot{\bf r}_i + m \boldsymbol{\omega} \times {\bf r}_i - {\bf B}_i)\vspace{1.0ex}\\
& + \dot{\gamma} (\dot{\gamma} \ineta + \boldsymbol{\omega}\cdot {\bf C}_{\gamma} + S_{\gamma})
\end{split}
\end{equation}
Rearranging the terms in this equation leads to:
\begin{equation}
\begin{split}
E & = \boldsymbol{\omega} \cdot {\bf I} \cdot \boldsymbol{\omega} + 2 \boldsymbol{\omega} \cdot \sum_i m ({\bf r}_i \times \dot{\bf r}_i)
+ \sum_i m \dot{\bf r}_i^2\vspace{1.0ex}\\
& + \dot{\gamma}^2 \ineta + 2 \dot{\gamma} \boldsymbol{\omega} \cdot {\bf C}_{\gamma} + \boldsymbol{\omega} \cdot {\bf S}_{\rm R}\vspace{1.0ex}\\
&- \sum_i {\bf B}_i \cdot (\dot{\bf r}_i +  \boldsymbol{\omega} \times {\bf r}_i)
+ \dot{\gamma} S_{\gamma}\vspace{1.0ex}\\
&= 2 \mathcal{L} + 2 V + \sum_i {\bf B}_i \cdot (\dot{\bf r}_i +  \boldsymbol{\omega} \times {\bf r}_i)\vspace{1.0ex}\\
& - \sum_i (\mu_{\rm B}^2/mc^2) ({\bf S}_i \times \boldsymbol{\mathcal{E}}_i)^2
 - \boldsymbol{\omega} \cdot {\bf S}_{\rm R}  - \dot{\gamma} S_{\gamma}\vspace{1.0ex}\\
\end{split}
\label{elagrangian}
\end{equation}
where $\mathcal{L}$ is the Lagrangian defined in Eq.~(2) of the \pap.
In agreement with Eq.~(13) of Curl~\cite{curl65} and Eq.~\eqref{edef}, the
Hamiltonian is:
\begin{equation}
H=E - \mathcal{L}
\label{hdef}
\end{equation}
Replacing $E$ in this equation by the half sum of the
results from Eqs.~\eqref{eham} and \eqref{elagrangian}
leads to $E=H+\mathcal{L}$, because several terms cancel.
Equation~\eqref{hdef} confirms that the operator in Eq.~(4) of the
\pap\ is indeed the Hamiltonian.
 
\subsection{\thesubsection~Quantum mechanical Hamiltonian}\stepcounter{subsection}
The exact quantum-mechanical Hamiltonian of a polyatomic
molecule is derived in Chapter~11 of Wilson, Decius, and
Cross' book~\cite{wilson_decius_cross}.  Their results are
used here considering $3 N_e$ vibrational degrees of freedom
corresponding to the 3 Cartesian coordinates of the electrons
and an additional degree of freedom corresponding to the torsional
motion of the methyl group. The quantum-mechanical Hamiltonian
is given in Eq.~(10) of Chapter~11. This equation involves
the tensor $\boldsymbol{\mu}$, which is the $3\times 3$
tensor defined for Eq.~(4) of the \pap, and its determinant
$\mu$. As stressed in the \pap, the tensor $\boldsymbol{\mu}$
is constant.  The determinant $\mu$ in Eq.~(10) of  Wilson,
Decius, and Cross~\cite{wilson_decius_cross} no longer needs
to be considered and this equation reduces to Eq.~(4) of the
\pap\ which is therefore the correct quantum-mechanical Hamiltonian of
the molecule.

\subsection{\thesubsection~The rotation-torsion Hamiltonian}\stepcounter{subsection}
The rotation-torsion Hamiltonian in Eq.~(5) of the
\pap\ is shown to be equivalent to the usual
RAM Hamiltonian as given in Eq.~(1) of Xu~{\em et al.}~\cite{xu08}.
The rotation-torsion Hamiltonian in Eq.~(5) of the
\pap\ can be rewritten as follows:
\begin{equation}
\begin{split}
H_{\rm RT} & ={\textstyle\frac{1}{2}}{\bf N} \cdot\boldsymbol{\mu} \cdot {\bf N}\vspace{1.0ex}\\
& -{\textstyle\frac{1}{2}}({\bf N} \cdot\boldsymbol{\mu} \cdot \boldsymbol{\lambda} + \boldsymbol{\lambda} \cdot\boldsymbol{\mu} \cdot {\bf N} ) p_{\gamma}\vspace{1.0ex}\\
& +{\textstyle\frac{1}{2}}( \boldsymbol{\lambda} \cdot\boldsymbol{\mu} \cdot \boldsymbol{\lambda} +  \frac{1}{I^{\alpha}})p_{\gamma}^2 + V_{\rm T}(\gamma)
\end{split}
\label{hrt}
\end{equation}
As in Xu~{\em et al.}~\cite{xu08}, the molecule fixed axis system is attached
to the molecule so that the two carbon and the oxygen atoms are located in the
$xz$ plane. The axis of internal rotation being also in that plane, we have $\lambda_y=0$.
The second term in Eq.~\eqref{hrt} reduces to:
\begin{equation}
-{\textstyle\frac{1}{2}}({\bf N} \cdot\boldsymbol{\mu} \cdot \boldsymbol{\lambda} + \boldsymbol{\lambda} \cdot\boldsymbol{\mu} \cdot {\bf N} ) p_{\gamma}=
 - p_{\gamma} {\bf N} \cdot {\bf v}
\label{lbddef}
\end{equation}
where ${\bf v}$ is the vector $\boldsymbol{\mu} \cdot
\boldsymbol{\lambda}$.  Due to the way the molecule fixed axis
system is attached to the molecule, we have $v_y=0$. Also,
it is possible to choose a new molecule fixed axis system,
differing from the original one by a rotation about the
$y$ axis, such that $v_x$ is also zero. Equation~\eqref{lbddef}
reduces then to $-v_z N_z p_{\gamma}$. The rotation-torsion
Hamiltonian in Eq.~\eqref{hrt} can then be rewritten as follows:
\begin{equation}
\begin{split}
H_{\rm RT} &= F(p_{\gamma} - \rho N_z)^2 + A N_z^2 + B N_x^2 + C N_y^2\vspace{1.0ex}\\
&+D\{N_x,N_z\} + V_{\rm T}(\gamma)
\end{split}
\label{HRAM}
\end{equation}
where:
\begin{equation}
\left\{\begin{array}{l@{\hspace*{3\tabcolsep}}l}
\displaystyle
F = {\textstyle\frac{1}{2}}( \lambda_z v_z +  1/ I^{\alpha}) & \rho = {\textstyle\frac{1}{2}}v_z / F\vspace{1.1ex}\\
\displaystyle
A = {\textstyle\frac{1}{2}}\mu_{zz} - F \rho^2 & B = {\textstyle\frac{1}{2}}\mu_{xx}\vspace{1.1ex}\\
\displaystyle
C = {\textstyle\frac{1}{2}}\mu_{yy} & D = {\textstyle\frac{1}{2}}\mu_{xz}
\end{array}\right.
\end{equation}
The rotation-torsion Hamiltonian in Eq.~\eqref{HRAM} clearly has the correct expression.

\subsection{\thesubsection~Variations of the fine splitting with $k$}\stepcounter{subsection}
The torsional Hamiltonian $H_{\rm T}$ of Eq.~(13) of the
\pap\ is diagonalized using the free internal rotation
basis set functions $|m\rangle= \exp(i m \gamma) /\sqrt{2\pi}$.
The diagonal matrix elements of $H_{\rm T}$ are the
following:
\begin{equation}
\begin{split}
\langle m | H_{\rm T} | m \rangle &= F( m - \rho k)^2 + e_z \langle S_z \rangle m \vspace{1.0ex}\\
&+ e_z \langle S_z \rangle k/2 + V_3/2
\end{split}
\label{htmatd}
\end{equation}
The nondiagonal matrix elements of $H_{\rm T}$ are:
\begin{equation}
\langle m | H_{\rm T} | m'\rangle = \left\{\begin{array}{ll}
-V_3/4 & \mbox{if $|m - m'| = 3$}\vspace{1.0ex}\\
0 & \mbox{otherwise}\end{array}\right.
\label{diaghtor}
\end{equation}
In Eq.~\eqref{htmatd}, $\langle S_z \rangle$ denotes
the diagonal matrix elements of $S_z$ between two Hund's case $b$ rotation-spin
functions:
\begin{equation}
\langle NkSJ | S_z | NkSJ\rangle = \left\{
\begin{array}{ll}
\displaystyle
+\frac{k}{2(N+1)} & \mbox{for F$_1$}\vspace{1.0ex}\\
\displaystyle
-\frac{k}{2N} & \mbox{for F$_2$}\end{array}\right.
\label{ndiaghtor}
\end{equation}
Numerical diagonalization of $H_T$ with $m$ in
Eqs.~\eqref{diaghtor} and \eqref{ndiaghtor}
having the value appropriate for the $C_3$ symmetries
leads to the electron spin-torsion energies plotted
in Fig.~(2) of the \pap.\vfill

\bibliography{gbib,ch3co,supp}